\documentstyle[epsfig]{mn}
\begin{document}

\input BoxedEPS.tex
\newcommand{\aip}{{\small ${\cal AIPS}$}}
\newcommand{\gtsim}{\mbox{{\raisebox{-0.4ex}{$\stackrel{>}{{\scriptstyle\sim}}
$}}}}
\newcommand{\ltsim}{\mbox{{\raisebox{-0.4ex}{$\stackrel{<}{{\scriptstyle\sim}}
$}}}}
\newcommand{\s}{$\stackrel{\rm s}{.}$}
\newcommand{\h}{$^{\rm h}$}
\newcommand{\m}{$^{\rm m}$}
\newcommand{\pp}{$\stackrel{\prime\prime}{.}$}
\newcommand{\de}{$^{\circ}$}
\newcommand{\p}{$^{\prime}$}
\newcommand{\arc}{$^{\prime\prime}$}
\newcommand{\marc}{^{\prime\prime}}
\newcommand{\rs}{{\em $r_s$}}
\newcommand{\DPM}{{\em DPM}}
\newcommand{\alf}{{\displaystyle\biggl({\nu_{\rm h} \over \nu_{\rm l}}\biggr)^{\alpha}} }

\newcommand{\figstart}[1]
    { \begin{figure}[htb]
      \begin{picture}(0,#1) }
\newcommand{\figend}[4]
    { \end{picture}
      \special{#1}
      \caption[#2]{#3}
      \label{#4}
      \end{figure} }
\newcommand{\fig}[5]
    { \figstart{#1}
      \figend{#2}{#3}{#4}{#5} }
\newcommand{\bHS}{\beta_{\mbox{\scriptsize HS}}}
\newcommand{\bBF}{\beta_{\mbox{\scriptsize BF}}}
\newcommand{\nT}{\nu_{\mbox{\scriptsize T}}}
\newcommand{\et}{E_{\mbox{\scriptsize T}}}
\newcommand{\nTn}{\nu_{\mbox{\scriptsize Tn}}}
\newcommand{\nTf}{\nu_{\mbox{\scriptsize Tf}}}
\newcommand{\tn}{\tau_{x\mbox{\scriptsize n}}}
\newcommand{\tf}{\tau_{x\mbox{\scriptsize f}}}
\newcommand{\xn}{x_{\mbox{\scriptsize n}}}
\newcommand{\xf}{x_{\mbox{\scriptsize f}}}
\newcommand{\yn}{y_{\mbox{\scriptsize n}}}
\newcommand{\yf}{y_{\mbox{\scriptsize f}}}
\newcommand{\lln}{l_{\mbox{\scriptsize n}}}
\newcommand{\llf}{l_{\mbox{\scriptsize f}}}
\newcommand{\Dn}{f(\Delta_{\mbox{\scriptsize n}})}
\newcommand{\Df}{f(\Delta_{\mbox{\scriptsize f}})}
\newcommand{\B}{\mbox{$B$}}
\newcommand{\Bo}{\mbox{$B$}_{0}}


\title[ELAIS Final Catalogue]{The European Large Area {\em ISO} Survey (ELAIS): The Final Band-merged Catalogue}

\author[Rowan-Robinson M. et al]
{M.Rowan-Robinson$^1$,
C.Lari$^{2}$,
I.Perez-Fournon$^{3}$, 
E.A.Gonzalez-Solares$^{4}$, 
\newauthor
F.La Franca$^{5}$,
M.Vaccari$^{6}$, 
S.Oliver$^{7}$, 
C.Gruppioni$^{8}$, 
P.Ciliegi$^{8}$,
\newauthor 
P.H\'eraudeau$^{9}$,
S.Serjeant$^{10}$,
A.Efstathiou$^{11}$,
T.Babbedge$^1$,
I.Matute$^{5}$,
\newauthor
F. Pozzi$^{8}$,
A.Franceschini$^{6}$,
P.Vaisanen$^{12,36}$,
\newauthor
A.Afonso-Luis$^{3}$,
D.M.~Alexander$^{4}$,
O.~Almaini$^{13}$,
A.C.Baker$^{19}$,
S.Basilakos$^{17}$,
\newauthor
M.~Barden$^{22}$,
C.del Burgo$^{33}$,
I.~Bellas-Velidis$^{17}$,
F.~Cabrera-Guerra$^{3}$,
\newauthor
R.~Carballo$^{18}$,
C.J.~Cesarsky$^{14}$,
D.L.~Clements$^{1}$,
H.~Crockett$^{1}$,
L.~Danese$^{20}$,
\newauthor
A.~Dapergolas$^{17}$,
B.~Drolias$^{1}$,
N.~Eaton$^{1}$,
E.~Egami$^{21}$,
D.~Elbaz$^{19}$,
D.~Fadda$^{10}$,
\newauthor
M.~Fox$^{1}$,
R.~Genzel$^{16}$,
P.~Goldschmidt$^{1}$,
J.I.Gonzalez-Serrano$^{15}$,
M.~Graham$^{1}$,
\newauthor
G.~L.~Granato$^{6}$,
E.Hatziminaoglou$^{3}$,
U.~Herbstmeier$^{22}$,
M.~Joshi$^{1}$,
E.~Kontizas$^{17}$,
\newauthor
M.~Kontizas$^{23}$,
J.K.~Kotilainen$^{24}$,
D.~Kunze$^{16}$,
A.~Lawrence$^{13}$,
D.~Lemke$^{22}$,
\newauthor
M.J.D.~Linden-V{\o}rnle$^{25,26}$,
R.G.~Mann$^{13}$,
I.~M{\'a}rquez$^{27}$,
J.~Masegosa$^{27}$,
\newauthor
R.G.~McMahon$^{4}$,
G.~Miley$^{28}$,
V.~Missoulis$^{1}$,
B.~Mobasher$^{29}$,
T.~Morel$^{35}$,
\newauthor
H.~N{\o}rgaard-Nielsen$^{26}$,
A.~Omont$^{31}$,
P.~Papadopoulos$^{28}$,
J-L.~Puget$^{31}$,
\newauthor
D.~Rigopoulou$^{34}$,
B.~Rocca-Volmerange$^{30}$,
N.Sedgwick$^{10}$,
L.~Silva$^{20}$,
T.~Sumner$^{1}$,
\newauthor
C.~Surace$^{1}$,
B.Vila-Vilaro,$^{21}$,
P.~van~der~Werf$^{28}$,
A.~Verma$^{16}$,
L.~Vigroux$^{19}$,
\newauthor
M.~Villar-Martin$^{30,37}$,
C.J.~Willott$^{32}$,
A.Carrami\~nana$^{38}$,
R.Mujica$^{38}$
\\
$^{1}$ Astrophysics Group, Blackett Laboratory, Imperial College, Prince Consort Rd, London SW7 2BZ,\\
$^{2}$ Istituto di Radioastronomia, Via P.Gobetti 101, Bologna 40129, Italy, $^{3}$ Instituto de Astrofisica de Canarias, C/ Via Lactea,\\
38200 La Laguna, S/C de Tenerife, Spain, $^{4}$ Institute of Astronomy, Madingley Road, Cambridge, CB3 0HA, $^{5}$ Dipartimento di Fisica,\\ 
Universita degli Studi ``Roma TRE'' Via della Vasca Navale 84, I-00146, Roma, Italy, $^{6}$ Dipartimento di Astronomia, Universita' di \\
Padova, Vicolo Osservatorio 5, I-35122 Padova, Italy, $^{7}$ Astronomy Centre, Department of Physics \& Astronomy, University of Sussex,\\ 
Brighton, BN1 9QJ, UK, $^{8}$ Osservatorio Astronomico di Bologna, via Ranzani 1, 40127 Bologna, Italy, $^{9}$ Kapteyn Astronomical\\
Institute, Postbus 800, 9700 AV Groningen, Netherlands, $^{10}$ Centre for Astrophysics and Planetary Science, School of Physical\\ 
Sciences, University of Kent, Canterbury, Kent CT2 7HR, UK, $^{11}$ Dept of Computer Science and Engineering, Cyprus College,\\ 
6 Diogenes St, Engomi, 1516 Nicosia, Cyprus, $^{12}$ European Southern Observatory, Casilla 19001, Santiago, Chile,\\ 
$^{13}$ Institute for Astronomy, University of Edinburgh, Royal Observatory, Blackford Hill, Edinburgh EH9 3HJ,\\
$^{14}$ ESO, Karl-Schwarzschild-Str 2, D-85748 Garching bei Munchen, Germany, $^{15}$  Instituto de F\'{\i}sica de Cantabria (Consejo\\
Superior de Investigaciones Cient\'{\i}ficas - Universidad de Cantabria, 39005 Santander, Spain,  $^{16}$  Max-Planck-Institut f\"{u}r\\ 
Extraterrestrische Physik, Postfach 1603, 85740 Garching, Germany, $^{17}$ National Observatory of Athens, Astronomical Institute,\\ 
P) Box 20048, GR-11810, Greece, $^{18}$ Dpto. de Matematica Aplicada, Universidad de Cantabria, 39005 Santander, Spain,\\
$^{19}$ CEA / SACLAY, 91191 Gif sur Yvette cedex, France, $^{20}$ SISSA, International School for Advanced Studies, Via Beirut 2-4,\\
 34014 Trieste, Italy, $^{21}$ Steward Observatory, University of Arizona, 933 North Cherry Avenue, Tucson, AZ 85721-0065\\
$^{22}$ Max-Planck-Institut f\"{u}r Astronomie, K\"{o}nigstuhl (MPIA) 17, D-69117, Heidelburg, Germany\\
$^{23}$ Section of Astrophysics, Astronomy \& Mechanics, Dept. of Physics, University of Athens, Panepistimiopolis, GR-15783,\\
 Zografos, Greece, $^{24}$ Tuorla Observatory, University of Turku, V\"ais\"al\"antie 20, FIN-21500 Piikki\"o, Finland\\
$^{25}$ Niels Bohr Institute for Astronomy, Physics and Geophysics, Astronomical Observatory, Juliane Maries Vej 30, DK--2100 \\
Copenhagen {\O}, Denmark, $^{26}$ Danish Space Research Institute, Juliane Maries Vej 30, DK--2100 Copenhagen {\O}, Denmark\\
$^{27}$ Instituto de Astrof\'{i}sica de Andaluc\'{i}a, CSIC, Apartado 3004,  E-18080 Granada, Spain, $^{28}$ Leiden Observatory,\\ 
P.O.Box 9513, NL-2300 RA Leiden, Netherlands, $^{29}$ Space  Telescope Science Institute, Baltimore, FA, USA,\\ 
$^{30}$ Institut d'Astrophysique de Paris, 98bis Boulevard Arago, F 75014 Paris, France, $^{31}$ Institut d'Astrophysique Spatiale\\
(IAS),  B\^{a}timent 121, Universit\'{e} Paris XI, 91405 Orsay cedex, France, $^{32}$ Herzberg Institute of Astrophysics, National\\ 
Research Council, 5071 West Saanich Rd, Victoria, BC V9E 2E7, Canada, $^{33}$ ESTEC, Keplerlaan 1, Postbus 299, 2200 AG Nordwijk,\\ 
Netherlands, $^{34}$ Physics Dept., University of Oxford, Denys Wilkinson Building, Keble Rd, Oxford OX1 3RH, UK,\\ 
$^{35}$ Istituto Nazionale di Astrofisica, Osservatorio Astronomico de Palermo G.S.Vaiana, Piazza del Parlamento 1, I-90134, Palermo,\\
Italy, $^{36}$ Observatory, Tahtitorninmaki, FIN-00014 University of Helsinki, Finland, $^{37}$ Dept. of Physical Sciences, University of \\
Hertfordshire, College Lane, Hatfield, Herts, AL10 9AB, UK, $^{38}$ INAOE, Luis Enrrique Erro 1, Tonantzintla, Puebla, Mexico\\
}

\maketitle
\begin{abstract}
We present the final band-merged ELAIS Catalogue at 6.7, 15, 90, and 175 $\mu$m, and the associated data at 
U,g',r',i',Z,J,H,K,  and 20cm.
The origin of the survey, infrared and radio observations, data-reduction and optical identifications 
are briefly reviewed, and a summary of the area covered, and completeness limit for each infrared band is given.
A detailed discussion of the band-merging and optical association strategy is given.
The total Catalogue consists of 3762 sources.
23 $\%$ of the 15 $\mu$m sources and 75 $\%$ of the 6.7 $\mu$m sources are stars. 
For extragalactic sources observed in 3 or more infrared bands,
colour-colour diagrams are presented and discussed in terms of the contributing infrared
populations.  Spectral energy distributions (seds) are shown for selected sources and compared with cirrus,
M82 and Arp220 starburst, and AGN dust torus models.  

Spectroscopic redshifts are tabulated, where available.  For the N1 and N2 areas, the INT ugriz 
Wide Field Survey permits photometric redshifts to be estimated for galaxies and quasars.  These agree
well with the spectroscopic redshifts, within the uncertainty of the photometric method ($\sim 10 \%$ in
(1+z) for galaxies).  The redshift distribution is given for selected ELAIS bands and colour-redshift diagrams 
are discussed.

There is a high proportion of ultraluminous infrared galaxies ($lg_{10}$ of 1-1000 $\mu$m luminosity $L_{ir} > 12.22$)
in the ELAIS Catalogue ($ 14 \%$ of 15 $\mu$m galaxies with known z),
many with Arp220-like seds. 10 $\%$ of the 15 $\mu$m sources are genuine optically blank fields to r' = 24: 
these must have very high
infrared-to-optical ratios and probably have z $>$ 0.6, so are high luminosity dusty starbursts or
Type 2 AGN.  9 hyperluminous infrared galaxies ($L_{ir} > 13.22$) and 9 EROs (r-K $>$ 6) are found
in the survey.  The latter are interpreted as ultraluminous dusty infrared galaxies at $z \sim 1$.
The large numbers of ultraluminous galaxies imply very strong
evolution in the star-formation rate between z = 0 and 1.
There is also a surprisingly large population of luminous ($L_{ir} > 11.5$), cool (cirrus-type seds)
galaxies, with $L_{ir}-L_{opt} > 0$, implying $A_V > 1$.

\end{abstract}
\begin{keywords}
infrared: galaxies - galaxies: evolution - star:formation - galaxies: starburst - 
cosmology: observations
\end{keywords}


\section{Introduction}
The European Large Area {\em ISO} Survey (ELAIS) was originally proposed in response to the first
Infrared Space Observatory ({\em ISO}) call for open time proposals in 1995 by M.Rowan-Robinson and 11
co-investigators from 9 institutions (Rowan-Robinson et al 1999).  Subsequently the collaboration 
has grown to a total
of 77 investigators from 32 European institutions.  The original concept was for a survey
of 12 sq deg of sky at wavelengths of 15 and 90 $\mu$m. Subsequent awards of observing time
allowed the survey to be extended to 6.7 $\mu$m and (in collaboration with the FIRBACK team
led by J.-L.Puget) 175 $\mu$m.  The ELAIS areas were also surveyed at 20 cm with the VLA and the
AT.

The survey goals, selection of survey areas, and details of the {\em ISO} observations, were described
by Oliver et al (2000), and a preliminary analysis of the 6.7 and 15 $\mu$m data was reported
by Serjeant et al (2000) and of the 90 $\mu$m data by Efstathiou et al (2000).  A new method
of reduction of the 15 $\mu$m data, which incorporates a physical models of cosmic ray and
transient effects, was given by Lari et al (2001), and a first application of this to the
ELAIS S1 area was described by Gruppioni et al (2001).  The final 15 $\mu$m reduction of the S1 
area is reported by Lari et al (2004), of the S2
area by Pozzi et al (2003) and of the N1, N2 and N3 areas by Vaccari et al (2004)
(see below for explanation of survey areas).  The final reduction of the 90 $\mu$m data is
reported by H\'eraudeau
 et al (2004) and of the FIRBACK-ELAIS 175 $\mu$m
data by Dole et al (2002).  A future paper by Rodighiero et al (2004, in preparation) will report
results of applying the Lari method to the ELAIS 90 $\mu$m data. The reduction and analysis of the 20 cm data was described
by Ciliegi et al (1999) for N1, N2 and N3, and by Gruppioni et al (1999) for S1 and S2.

Associated with the ELAIS survey there has also been an extensive program of ground-based optical
and near-infrared imaging and spectroscopy.  The optical and spectroscopic follow-up of the
S1 area has been presented by La Franca et al (2004), of the S2 area by Pozzi et al (2003),
and of the N1 and N2 areas by H\'eraudeau et al (1999), Vaisanen et al (2002), Gonzalez-Solares et al (2004),
Perez-Fournon et al (2004), Serjeant et al (2004), Verma et al (2004) and Afonso-Luis et al (2004).

Deeper surveys at 6.7 and 15 $\mu$m were also carried out by a subset of the ELAIS consortium in
HDF-N (Serjeant et al 1997, Goldschmidt et al 1997, Oliver et at 1997, Mann et al 1997, Rowan-Robinson et al 1997)
and in HDF-S (Oliver et al 2002, Mann et al 2002).

X-ray surveys have also been carried out in several ELAIS areas.  Alexander et al (2001) reported
Beppo-Sax observations over a large fraction of S1.  Manners et al (2003) have reported
Chandra
observations in the central regions of N1 and N2, and some interpretation of these observations
has been given by Willott et al (2003).  ROSAT data were compared with the ELAIS Preliminary Catalogue
by Basilakos et al (2002).

The present paper reviews the parameters of the ELAIS survey, gives a detailed account of the merging
of the different individual wavelength surveys to generate a final catalogue, and discusses
the populations present in the survey through colour-colour and colour-redshift diagrams,
and spectral energy distributions.

\section{Quality of the constituent infrared catalogues}
The separate wavebands making up the ELAIS survey each comprise an independent survey and
are discussed in separate papers.  The present paper does not replace these analyses but 
pulls together those aspects which emerge from band-merging the surveys.  For detailed
discussion and analysis of the ELAIS sources, it is essential to refer back to these
analyses of the individual constituent surveys.  Table 1 defines the survey areas,
wavelengths, characteristic depth, and source-densities.

\begin{table*}
\vspace{4cm}
\caption{Summary of survey wavelengths, areas and numbers of sources, characteristic depth,
source-densities}
\begin{tabular}{lllllll}
Name & RA & Dec & 6.7 & 15 & 90 & 175 $\mu$m \\
 & & & & & & \\
N1 & 16$^h$10$^m$01$^s$ & +54$^o$30'36" & & 2.67/490 & 2.56/151 & 2/103 (sq deg/no. of sources)\\
N2 & 16 36 58 & +41 15 43 & 2.67/628 & 2.67/566 & 2.67/174 & 1/55 \\
N3 & 14 29 06 & +33 06 00 & 1.32/189 & 0.88/131 & 1.76/119 & \\
S1 & 00 34 44 & -43 28 12 & 1.76/304 & 3.96/317 & 3.96/226 & \\
S2 & 05 02 24 & -30 35 55 & 0.12/40 & 0.12/43 & 0.11/5 & \\
 & & & & & &\\
& total area/number &  & 5.86/1161 & 10.3/1546 & 11.06/674 & 3.0/158 sq deg/no. of sources \\
 & & & & & & \\
& characteristic depth & & 1.0 & 0.7 & 70 & 223 mJy \\
& source-density & & 200 & 150 & 61 & 53 per sq deg \\
\end{tabular}
\end{table*}

The calibration, completeness and reliability of the ELAIS 15 $\mu m$ survey final analysis has been discussed by
Lari et al (2001, 2003), Gruppioni et al (2001) and Vaccari et al (2003).  
In the Lari method (Lari et al 2003), the ISOCAM Handbook (Blommaert et al 1998) conversion factor from {\em ISO} instrumental units (ADU)
to mJy of 1.96 is used, with full modelling of the detector stabilization.  To achieve consistency with the IRAS Faint
Source Catalog 12 $\mu$m calibration a multiplicative factor of 1.097 is applied to 15 $\mu$ fluxes, based on the
analysis by Aussel and Alexander (2003, in preparation) of near infrared, ISO and IRAS fluxes for stars (Vaccari et al 2003).
The IRAS 12 $\mu$m calibration has been tested by Aussel and Alexander for over 4000 stars detected by IRAS and by 2MASS.
By contrast, the Preliminary Analysis (PA) of Serjeant et al (2000) used a direct conversion of 1.75 ADU/mJy,
based on a comparison of ISO and IRAS fluxes for stars,
without correction for detector stabilization.  Subsequently Vaisanen et al (2002) used J and K observations
of ISO stars to derive a corrected conversion factor for the PA Catalogue of 1.05 ADU/mJy.
 Figure 1 shows a plot of the final 15 $\mu m$ flux, $S_{15}$, versus the PA flux (in ADU) (Babbedge 2004),
which supports the conclusions of Vaisanen et al (2002).
The Final Analysis Catalogue goes significantly deeper than the PA Catalogue, but almost all PA sources
are found to be real, though much of the scatter in Fig 1 must be attributed to lower accuracy of the PA fluxes.

It would also have been desirable to apply the Lari method to the 6.7 $\mu$m survey, but we have not been able to
put together the very considerable resources needed to do this.  As we will see in section 3.3 75 $\%$ of the
6.7 $\mu$m sources are stars so the scientific returns for the present, predominantly extragalactic, study did
not seem great.  We have relied on the analysis of Vaisanen et al (2002) which confirmed both the reality of
the 6.7 $\mu$m Preliminary Analysis sources, and their flux calibration.  The SIRTF-SWIRE survey (Lonsdale et al
2003) will allow comparison of the 6.7 $\mu$m ISO fluxes with SIRTF 8 $\mu$m fluxes, for sources in N2 and S1.
The 6.7 $\mu$m PA Catalogue has duplicate entries where different rasters observed 
the same part of the sky and only the entry with lowest noise was included.  

The calibration, completeness and reliability of the ELAIS 90 $\mu m$ survey final analysis has been
discussed by H\'eraudeau et al (2004).  Calibration of 90 $\mu$m photometry is based on 
a direct comparison of standard stars.
Figure 2 shows a plot of the final 90 $\mu m$ flux, $S_{90}$,
versus the PA flux (Babbedge 2004).  Again, most of the PA sources are confirmed.  The substantial scatter can be
mainly attributed to the limitations of the PA processing. 

The 175 $\mu m$ data analysis and identifications have been discussed by Dole et al (2002).  A 5-$\sigma$
sensitivity of 223 mJy was achieved.
 
\begin{figure}
\epsfig{file=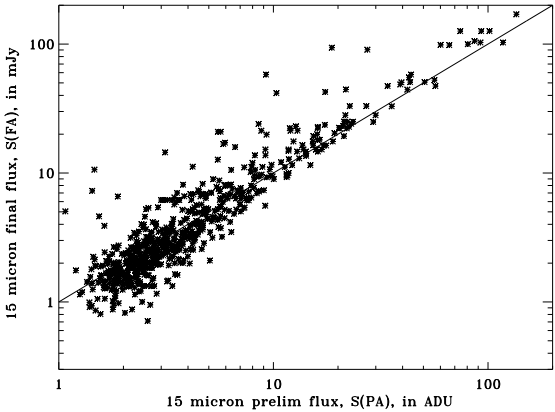,angle=90,width=8cm}
\caption{Comparison of Final Analysis 15 $\mu m$ flux (mJy) (Lari et al 2004, Vaccari et al 2004) with
Preliminary Analysis 15 $\mu m$ fluxes (ADU) (Serjeant et al 2000). The line corresponds to equal fluxes. 
}
\end{figure}

\begin{figure}
\epsfig{file=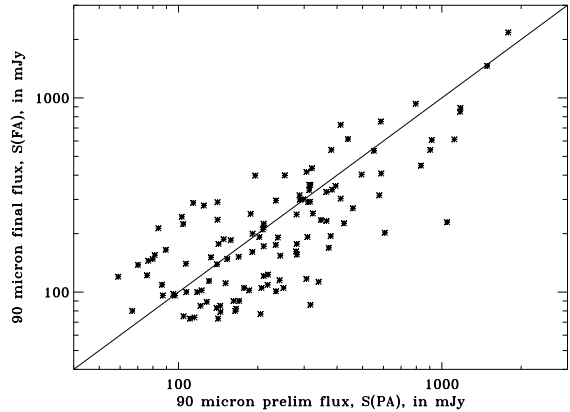,angle=90,width=8cm}
\caption{Comparison of Final Analysis 90 $\mu m$ flux (mJy) (H\'eraudeau
et al. 2003) with
Preliminary Analysis 90 $\mu m$ fluxes (Efstathiou et al 2000). 
}
\end{figure}

\section{Final Band-merged Catalogue}

\subsection{Band-merging}
We have proceeded with the band-merging of the multi-wavelength ELAIS catalogues in a sequential way,
taking into account the different positional accuracies of the component catalogues.  The
1-$\sigma$ positional uncertainties at 15 $\mu$m and 20cm have been estimated to be $\sim$ 1" (Lari et al
2004, Vaccari et al 2004, Gonzalez-Solares et al 2004, Ciliegi et al 1999, Gruppioni et al 1999),
so sources at these wavelengths can be very reliably associated with optical counterparts, down
to at least r' $\sim$ 23 mag.  We therefore first separately identify the 15 $\mu$m and
20 cm catalogues with optical sources, using the likelihood method of Mann et al (1997),
as discussed by Gonzalez-Solares et al (2004).   
The latter have described detailed simulations of this association process, which show that the
probability of spurious associations is $< 5 \%$ for r' $<$ 20, rising to 20 $\%$ by r' = 24.  The 15
$\mu$m and 20 cm sources are then merged on the basis of their optical positions, using a search
radius of 2".  If the optical positions agree to this accuracy, the radio flux and error are added 
to the 15 $\mu$m catalog entry.  The percentages of 15 $\mu$m and 20 cm sources which found matches 
with the other
wavelength were 8$\%$ and 11$\%$ respectively.  A flag is set for sources for which there is more
than one candidate identification (see below): the selected association is the one with the
highest likelihood (Gonzalez-Solares et al 2004).
 
Radio sources which did not find a 15 $\mu$m match are then interleaved with the matched sources 
to give an RA-ordered list.  Matched sources are given the 15 $\mu$m source name. 

We next matched the 6.7 $\mu$m PA sources with the combined 15 $\mu$m-20 cm list, using a
search radius of 5".  The larger search radius is required because of the poorer astrometry of
the PA Catalogue, which is based on a single raster, with pixel size 3".
30 $\%$ of 6.7 $\mu$m sources found a match with a Catalogue source.

Non-matched 6.7 $\mu$m sources were associated with optical counterparts where possible, 
and then interleaved to generate a
combined 6.7-15 $\mu$m-20 cm source-list.  The positions of all sources in this list which did not 
have 20 cm fluxes were examined in the ELAIS 20 cm data to see whether a source in the 3-5 $\sigma$ range
might be present, and also in the VLA FIRST survey for areas not surveyed by ELAIS, 
and if so the flux was added to the Catalogue. Otherwise 3-$\sigma$ radio limits are given as negative
entries in the Catalogue.

The 90 and 175 $\mu$m sources were then associated with this combined list using search radii of
30 and 60" respectively.  Where a 90 or 175 $\mu$m source is matched with more than one Catalogue entry,
the less probable associations are flagged and are not used in subsequent discussions of infrared
colours or spctral energy distributions, i.e. all the flux is assigned to the most likely match. 
The non-matched 5-$\sigma$ 90 and 175 $\mu$m sources are retained as a separate supplementary catalogue, because
of their much poorer positions.  They were searched for optical counterparts of
high likelihood using search radii of 45 and 90". 
Because of the large positional uncertainties at 90 and 175 $\mu$m, 
optical counterparts have to be reasonably bright to have a high
likelihood of being the correct association
($r < 19.0$).
20 $\%$ of 90 $\mu$m sources and 1 $\%$ of 175 $\mu$ m sources, brighter than  5 $\sigma$,
found matches neither with the combined 6-7-15 $\mu$m-20 cm source-list nor with optical
counterparts.  Most of these may be associated with galaxies fainter than r' = 19.  However we
can not rule out the possibility that some of the unassociated 90 $\mu$m sources may be spurious.


Finally the combined 6.7-15-90-175 $\mu$m-20 cm Catalogue is searched for pairs within 5" and each of these
is examined individually.  In most, but not all, cases these pairs are believed to be the same object, split
in two by the optical association process, and in these case the sources have been merged, with an appropriate
flag set in the Catalogue.

\subsection{Catalogue Description}

The Final Band-merged Catalogue is given at 
http://astro.imperial.ac.uk/Elais/ .
The Catalogue entries are : source name (for merged sources, in order of preference 15 $\mu$m, 
20 cm, 6.7, 90, 175 $\mu$m ), source position (same order of preference for merged sources),  
20 cm flux and error, 175 $\mu$m flux, error, S/N, positional offset, 90 $\mu$m flux, error, S/N, positional offset,
15 $\mu$m
 flux and S/N, 6.7 $\mu$m flux and error, 6.7 $\mu$m reliability flag (2 = high reliability, 3 = medium
reliability), flag 1 (see below), flag 2, J mag and error, H mag and error, K mag and error, flag 3,
position of optical
association, WFS U, g', r', i', Z magnitudes and errors, WFS star/galaxy flags (-1 for stellar image, 1 for extended image), 
Sextractor r' mag and error,
Sextractor star/galaxy classification (0.0 for galaxy, $>$ 0.7 for stellar image), positional offset of optical ID 
(total, RA, Dec), probability of
optical association (threshold 0.8, see eqn (5) of Gonzalez-Solares et al (2004) for definition), 
reliability of optical association (see eqn (6) of Gonzalez-Solares et al (2004) for definition),
photometric spectral energy distribution (sed) type, $n_{typ}$, and redshift ($\lg_{10}(1+z_{phot})$, 
derived assuming $A_V$ = 0), photometric sed type and redshift and $A_V$ (free fit for $A_V$), spectroscopic 
redshift $z_{spect}$, flag 4, flag 5, $z_{best}$ = $z_{spect}$ if available, = $z_{phot}$ otherwise,
bolometric optical luminosity $L_{opt}$, ir sed type (1 = cirr, 2 = M82-sb, 3 = A220-sb, 4 = AGN dust torus,
5 = cooler cirrus, 6 = 2+4), bolometric (1-1000 $\mu$m) infrared luminosity $L_{ir}$,
AGN dust torus luminosity $L_{tor}$ (if 15 $\mu$m emission is interpreted as
due to dust torus emission).

where flag 1 = 1 if radio flux force-merged (15 $\mu m$ and radio positions within 5"),
= 2 if 6.7 $\mu m$ flux force-merged,
= 3 if 90 $\mu m$ association is not the most likely, where there are multiple associations,
= 4 if 175 $\mu m$ association is not the most likely, where there are multiple associations,
= 5 if second most probable optical association for 15 $\mu m$ has been preferred on basis of radio position,
= 6 if source has an {\em IRAS} association,
= 7 if 1 and 6 both set,
= 8 if 3 and 4 both set,
= 9 if 1 and 8 both set, or 4 and 5 both set;

flag 2 = 1 if source falls in gaps between WFS chips, so no optical data,
= 2 if source falls near edge of WFS chip, so photometry may be unreliable,
= 3 if source has multiple optical counterparts,
= 4 if source is blank in optical (ie no optical counterpart with probability of association $>$ 0.7,
within specified search radius),
= 6 if there is an association in NED (non-ELAIS, non-2MASS, non-IRAS),
= 7 if association is bright star,
= 8 if flags 1 and 7 set,
= 9 if flags 3 and 7 set;

flag 3 = 1 if J,H,K magnitudes from 2MASS,
= 2 if J, K magnitudes from Vaisanen et al (2002),
= 3 if K magnitudes from Rigopoulou et al (2004, in preparation),
= 4 if K magnitude from Pozzi et al (2003a)
= 5 if K magnitude from Sajina et al (2003);

flag 4, $n_{ztyp}$ (spectroscopic classification), = 1 (spiral) galaxy,
= 2  emission-line, starburst,
= 3  absorption line, early type,
= 4  AGN,
= 5  Sy1,
= 6  Sy2,
= 7  star, =8 liner;

flag 5, $n_{zref}$ = 1  Perez-Fournon et al (2003),
= 2  Serjeant et al (2003),
= 3  La Franca et al (2003),
= 4  Pozzi et al (2003a),
= 5  NED, = 6 Morel et al (2001), = 7 SLOAN Survey, = 8 Willott et al (2003), = 9 Chapman et al (2002), Sajina et al
(2003).

For extended objects in N1, N2, the WFS magnitudes refer to the flux within a fixed aperture: they should give the correct colours
of the objects, as required by the photometric redshift code.  For integrated magnitudes the Sextractor r' magnitude
should be used (and other WFS bands can be corrected by the difference between the WFS and Sextractor r' magnitudes).
Because of the effects of saturation, colours can only be trusted if the Sextractor r' magnitude is $>$ 15 (this mainly
affects stars in the Catalogue).  In S1, the B magnitudes are derived from APM magnitudes and are very uncertain (also
possibly too faint on average by $\sim$ 0.5 mag.).

There are 3523 sources in the 6.7-15$\mu$m-20cm Catalogue, of which 1636 are 15 $\mu$m sources, 1136 are 20 cm (non 15 $\mu$m) sources, 
741 are 6.7 $\mu$m (non 15 $\mu$m, non 20 cm) sources, and
239 sources are in the supplementary 90-175$\mu$m catalogue.  The numbers of 15 $\mu$m sources with redshifts
are 204 in S1 (199 spectroscopic), 31 in S2 (22 spectroscopic), 10 in N3 (all spectroscopic), 309 in N1 (109 spectroscopic) and
355 in N2 (167 spectroscopic), 909 15 $\mu$m  redshifts in all, and 1210 redshifts in the whole Catalogue (see below for discussion of photometric redshifts).  97 $\%$ of 
15 $\mu$m sources in N1 and N2 are accounted for as either stars, extragalactic sources with redshifts, or blank fields
(8 and 11 $\%$ respectively), and 93 $\%$ in S1 (the percentage of blank fields in S1 is higher, 20 $\%$, 
because of the shallower optical survey).
48 $\%$ of the radio (non 15 $\mu$m) sources are blank fields (to r' = 24).  91 $\%$ of the 6.7 $\mu$m (non 15 $\mu$m or radio) sources
in N1 are found to be stars.

\subsection{Associations}
We matched the whole Catalogue with the NASA Extragalactic Database
(NED) and  redshifts resulting from these associations are included in
our Catalogue.  We also specifically matched the Catalogue to the 2MASS
J,H,K catalogues, finding matches for 30 $\%$ of our sources (with a
higher success rate for the stars).
A search radius of 5" was used, and in the few cases where sources appeared in both the 2MASS Extended and Point-Source
Catalogues, the magnitudes from the Extended Catalogue were preferred.
Here we discuss some more specific results from
associations with known objects.

Table 2 lists the {\em ISO} and {\em IRAS} data for ELAIS
sources which are associated with {\em IRAS} Faint Source Catalog sources (FSC,
Moshir et al. 1992), with upper limits indicated as negative values.  Of the 39 associations in N1 and N2, 12 are
clearly stellar photospheres detected by {\em IRAS} at 12 $\mu$m (and
occasionally at 25 $\mu m$) and the rest are nearby normal galaxies.
For the galaxies detected by IRAS at 60 and 100 $\mu$m, we have interpolated to
estimate a 90 $\mu$m flux and calculated $log_{10} S(90)_{IRAS}/S(90)_{ISO}$.
The mean value for 19 sources is 0.129 $\pm$0.022, suggesting that there are unresolved calibration
issues between IRAS and ISO.  However it should be noted that there is a tendency
for the IRAS FSC to overestimate fluxes near the FSC threshold at 60 and 100 $\mu$m
(Moshir et al 1992).  Since all the IRAS sources in Table 2 have S(100) $<$ 3 Jy
(and most have S(100) $<$ 1 Jy), this would be sufficient to explain the
discrepancy noted above.

\begin{table*}
\caption{Matches with the IRAS Faint Source Catalog Version 2}
\begin{tabular}{lllllllllll}
IRAS name & 12 & 25 & 60 & 100 $\mu$m & IRAS-ISO & ELAIS name & 6.7 & 15 & 90 & 175 $\mu$m \\
 & mJy & & & & sepn (') ) & mJy & & & & \\
& & & & & & & & & &  \\
 F00279-4253 & -75.4  & -76.4  & 649.0  & 1440.0 & 0.70 & ELAISC15-J003022-423657 & -    & 25.21  & 849 & - \\
 F00302-4249 & -112.0 & -99.0  & 232.0  & -525.0 & 0.53 & ELAISC15-J003244-423313 & -    & 11.64  & 191 & - \\
 F00315-4421 & 178.0  & -93.8  & -123.0 & -376.0 & 0.43 & ELAISC15-J003402-440442 & -    & 109.34 & -     & - \\
 F00320-4342 & -65.8  & -103.0 & 309.0  & 685.0  & 0.25 & ELAISC15-J003429-432614 & 6.18 & 24.32  & 408 & - \\
 F00325-4313 & -96.6  & -75.5  & 322.0  & 906.0  & 0.13 & ELAISC15-J003458-425733 & -    & 15.83  & 534 & - \\
 F00341-4428 & -129.0 & -155.0 & 202.0  & -676.0 & 0.78 & ELAISC15-J003626-441140 & 5.80 & 13.61  & 290 & - \\ 
 F00353-4418 & -53.4  & -70.8  & 204.0  & -559.0 & 0.20 & ELAISC15-J003741-440226 & -    & 1.36   & 210 & - \\
 F00360-4355 & -124.0 & -128.0 & 374.0  & 1280.0 & 0.27 & ELAISC15-J003828-433848 & 10.14& 46.58  & 682 & - \\
 F00362-4416 & 135.0  & -149.0 & -166.0 & -279.0 & 0.12 & ELAISC15-J003836-440029 & -    & 27.48  & -     & - \\
& & & & & & & & & &  \\ 
 F14262+3328 &   -60.2 & -59.7  & 317.0  &  518.0 & 0.27 & ELAISC15-J142823.4+331513 & -    & 12.72 & 448 & - \\
 F14266+3336 &   -67.7 & -98.9  & 452.0  &  999.0 & 0.32 & ELAISC15-J142847.1+332315 & -    & 19.02 & 540 & - \\
 F14292+3318 &   143.0 & -116.0 & -105.0 & -316.0 & 0.18 & ELAISC15-J143123.5+330517 & 258.75  & 62.91 & -     & - \\
 F14292+3327 &   -68.4 &  94.9  & 694.0  &  983.0 & 0.13 & ELAISC15-J143125.3+331348 & 6.93 & 28.95 & 933 & - \\
 F14304+3341 &   -68.1 & -47.6  & 201.0  &  476.0 & 0.13 & ELAISC15-J143234.9+332833 & -    & 1.57  & 434 & - \\
& & & & & & & & & &  \\
 F16022+5450 &   -45.0 &  -65.7 & 159.0 & -614.0 & 0.70 & ELAISC15-J160322.8+544237 & - & 2.36 & 101 &   -\\
 F16029+5506 &   -66.7 & -67.0 & 296.0 & 507.0 &  0.09 & ELAISC15-J160408.4+545812 & - & 8.77 & 315 &   - \\
 F16046+5415 &   -83.2 & 78.8 & 604.0 & 798.0 &   0.25 & ELAISC15-J160552.5+540650 & - & 30.09 & 756 & 838  \\
 F16063+5405 &   -69.3 & -57.6 & 244.0 & -742.0 & 0.16 & ELAISC15-J160736.5+535731 & - &  19.15 &  230 & 597  \\
 F16070+5439 &    97.5 & -50.4 & -93.2 & -308.0 & 0.12 & ELAISC15-J160812.7+543141 & - &  53.25 &   -  &  -  \\
 F16083+5400 &   -48.3 & -42.6 & 184.0 & -720.0 & 0.11 & ELAISC15-J160934.7+535220 & - &   1.99 & 215 & 309  \\
 F16091+5357 &   207.0 & 69.9 & -102.0 & -419.0 & 0.09 & ELAISC15-J161019.3+534934 & - &  103.67 &   - &   - \\
 F16091+5447 &   576.0 & 146.0 & -91.2 7 & -244.0 & 0.02 & ELAISC15-J161017.6+543929 & - & 288.39 &   - &   -  \\
 F16145+5447 &   -35.7 & -64.4 & 161.0 & 475.0 & 0.33 &  ELAISC15-J161545.8+544019 & - &  20.70 & 303 &   - \\
 & & & & & & & & & &\\
 F16294+4115 &   -87.8 & -65.4 & 243.0 & 936.0 &  0.13 & ELAISC7-J163104+410913 &   2.96 &   - &  540 &   - \\
 F16298+4129 &   131.0 & -53.5 & -124.0 & -424.0 & 0.22 & ELAISC15-J163130.2+412330 & 255.16 &  66.21 &   - &   - \\
 F16323+4127 &   -74.9 & -67.8 & 383.0 & 743.0 &  0.23 & ELAISC15-J163401.8+412052 & 9.60 &  20.37 & 403 & 666 \\
 F16334+4116 &   -48.4 & -65.0 & 190.0 & -813.0 & 0.17 & ELAISC15-J163506.1+411038 & - &   7.98 & 251 & 346 \\
 F16337+4101 &   -66.4 & -73.7 & 224.0 & -843.0 & 0.17 & ELAISC15-J163525.1+405542 & 5.46 &  13.91 & 416 & 682 \\
 F16338+4138 &   147.0 & -70.2 & -104.0 & -322.0 & 0.12 & ELAISC15-J163531.1+413158 &  524.98 &  106.80 &   - &   - \\
 F16341+4053 &   103.0 & -80.2 & -88.2 & -369.0 & 0.24 & ELAISC15-J163549.0+404720 &  196.97 &  58.42 &   -  &  - \\
 F16341+4059 &   243.0 & -109.0 & -102.0 & -339.0 & 0.10 & ELAISC15-J163549.9+405317 & - & 125.86 &   - &   - \\
 F16344+4111 &   -71.9 & -67.6 & 351.0 & 897.0 &  0.06 & ELAISC15-J163608.1+410507 & 2.30 &   8.94 & 614 & 803 \\
 F16349+4038 &   -84.7 & -35.8 & 236.0 & 615.0 &  0.37 & ELAISC15-J163633.5+403245 & 6.15 &  16.52 & 291 &  - \\
 F16349+4034 &   116.0 & -50.7 & -98.7 & -441.0 & 0.04 & ELAISC15-J163637.3+402824 & - &  49.44 &   - &   - \\
 F16359+4058 &   -61.1 & 104.0 & 1220.0 & 2480.0 & 0.19 & ELAISC15-J163734.4+405208 & 18.37 &  53.56 & 1461 & 2377 \\
 F16365+4202 &   -69.6 & 84.4 & 460.0 & 1500.0 & 0.28 & ELAISC15-J163814.0+415620 & 6.14 &  50.15 & 611 &   - \\
 F16377+4150 &   -54.3 & -68.8 & 306.0 & 520.0 &  0.16 & ELAISC15-J163924.0+414442 & 2.55  &  8.13 &  299 &  - \\
 F16389+4146 &   249.0 & 81.2 & -86.2 & -500.0 & 0.06 & ELAISC15-J164033.9+414107 & 645.30  &  125.86 &  - &   - \\
 F16405+4113 &   -55.6 & -68.4 & 169.0 & -545.0 & 0.15 & ELAISC15-J164211.9+410816 & 1.80 &   8.32 & 288 &   - \\
\end{tabular}
\end{table*}


The ELAIS N1 field was also partially observed with the $H\alpha$
survey of Pascual et al. (2001). Since the infrared and $H\alpha$ both
trace star-formation it was tempting to see if there were any sources
in common.  Using the $H\alpha$, FIR, star-formation calibrations of
Cram et al. (1998) we were able to estimate a mean $H\alpha$ flux to FIR
flux. Then using a starburst SED we were able to estimate the expected
15$\mu$m flux for each of the Pascual et al. sources.  With one
exception the expected $15\mu$m fluxes all fell below  
the characteristic depth of the survey (0.7 mJy, Table 1).
The exception was the source in their field  a3 with ID 7227b at 
16~05~46.3 +54~39~11.74 with $m_I$ = 17.5 and $m_{\rm H\alpha}=17.0$.
We estimate that this source should have had a 15$\mu$m flux of 2.9
mJy. Neither this source nor any of the others were detected in our
ELAIS Catalogue. A more detailed analysis of the expected dispersion 
in the  $H\alpha$/FIR relation is required before we can assess
whether these non-detections suggest that the mean relation needs to
be revised.

The Canada-France-Hawaii Telescope blue grens quasar survey 
of Crampton et al. (1992) overlaps with the N2 field. 79 of their
candidates fall within the ELAIS boundaries.  11 of these candidates 
are detected by ELAIS and are listed in Table 3.

Stars can be recognized in the Catalogue through (i) having flag 2 set to 7, (ii) having flag 4
set to 7, (iii) having low ratios of $(S_{15}/S_r)$ or $(S_{6.7}/S_r)$, where $S_r$ is the r-band flux in mJy.
Almost all sources with $log_{10} (S_{15}/S_r) < -0.5 $, or $log_{10} (S_{6.7}/S_r) < -0.2$, are stars.
A total of 846 stars have been identified in this way in the Catalogue, comprising 22, 23 and
25 $\%$ of the 15 $\mu$m sources in N2, N1 and S1 respectively, and 75 $\%$ of 6.7 $\mu$m
sources.  In N1, 91 $\%$ of sources which are 6.7 $\mu$m (non-15 $\mu$m or 20 cm) sources
are stars.
The infrared fluxes are consistent with being photospheric emission in almost all cases.
The small number of stars where the ISO emission appears to be in excess of the photspheric
prediction deserve further detailed study, but this is beyond the scope of the present paper.

\begin{table*}
\caption{Quasar candidates from the blue grens survey of Crampton et
al. 1992 that are detected by ELAIS.  The name of the sources
in the Crampton et al catalogue and the ELAIS Catalogue are given.
This is followed by the separation of the two in arc seconds.  The
fourth column gives the spectral types estimated from the U, g,r,i,Z
INT data in our photometric redshift fitting.  The Q flag indicates
the confidence
of the quasar candidate in the Crampton et al. work from 1 (strong lines)
through to 4 (blue continuous spectrum). }
\begin{tabular}{llrrr}
Crampton et al. name & ELAIS name & Sep/arc sec &
opt sed type & Q \\
  1635.20+4124 & ELAISC15 J163652.7+411827 &      1.2 &       4&       2\\
  1635.30+4135 & ELAISC15 J163659.0+412928 &      2.5 &       6&       3\\
  1635.40+4136 & ELAISC15 J163702.2+413022 &      1.1 &       7&       4\\
  1635.70+4121 & ELAISC15 J163721.3+411503 &      0.7 &       8&       3\\
  1636.00+4149 & ELAISC15 J163739.3+414348 &      1.6 &       7&       2\\
  1636.50+4203 & ELAISC15 J163805.6+415740 &      1.4 &       3&       3\\
  1636.70+4133 & ELAISR163817+412730       &      1.3 &       7&       2\\
  1637.20+4217 & ELAISC15 J163847.5+421141 &      0.6 &       7&       4\\
  1637.60+4134 & ELAISC15 J163915.9+412834 &      1.0 &       6&       3\\
  1638.60+4126 & ELAISC15 J164016.0+412102 &      2.5 &       7&       1\\
  1638.70+4108 & ELAISC15 J164018.8+410254 &      1.2 &       8&
3\\
\end{tabular}
\end{table*}

\section{Radio-infrared-optical colour-colour diagrams}

For sources detected in 3 bands we can plot colour-colour diagrams and compare the results with predictions
of models.  Figs 3-8 show a selection of these, with predicted loci corresponding to the basic infrared 
templates used by Rowan-Robinson (2001): cirrus, M82 starburst, Arp220 starburst, AGN dust torus.

Figure 3 shows the 175/90/15 colour-colour diagram for N1 and N2, with loci for cirrus (C), M82 starburst (S), 
and Arp220
starburst (A), for redshifts from 0 to 3 (the position of the labels indicates the zero redshift end of the
loci).  The cross in the lower left-hand corner of this and subsequent diagrams indicates typical (median) error bars
(where these are larger than the plotted symbols).
The model loci loop around the diagram so without additional redshift information
it is hard to say much about the populations (or derive photometric redshifts from far infrared data).  However from
both spectroscopic and photometric redshifts (see section 5) we can
deduce that almost all these galaxies have  
$z<$ 0.5.
Most of the objects then lie reasonably close to the cirrus locus over this redshift range, so we deduce that this
bright subset consists of normal spirals with far infrared emission from interstellar dust bathed in the general
stellar radiation field.

Figure 4 shows the 90/15/6.7 colour-colour diagram for N2 and S1, with the same model loci.  Again there is ambiguity
about the populations involved, which is not resolved using the fact that most galaxies detected at 90 $\mu$m
have $z<$ 0.5.  An earlier version of this figure was discussed by Marquez et al (2002).

Figure 5 shows the 90/15/r-band colour-colour diagram for N1, with the same model loci.  Here the model loci are more
differentiated and we can deduce that all three templates are represented, with cirrus and M82 starburst
components accounting for over 75 $\%$ of the sources.

\begin{figure}
\epsfig{file=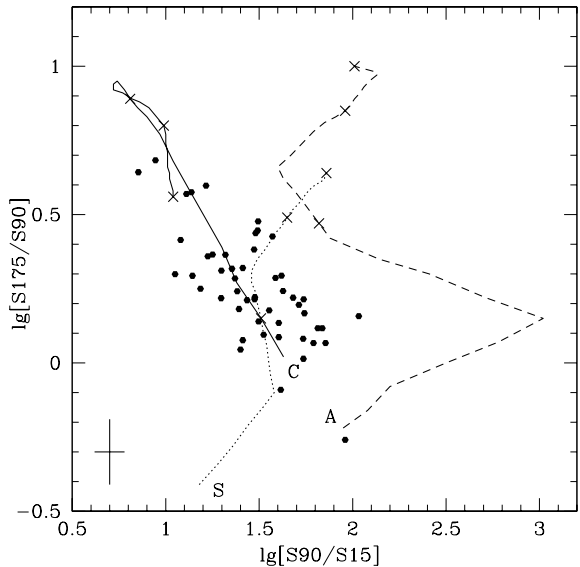,angle=0,width=8cm}
\caption{175-90-15 $\mu$m colour-colour diagram for sources in N1 and N2, with 
loci for cirrus (C),
starburst (S) and Arp220 (A) seds, from $z = $0 to 3 (labels denote the z = 0 end of the locus,
crosses mark z = 1, 2, 3).  Only sources detected
in all 3 bands are plotted: all are galaxies.
The cross in the lower left-hand corner of this and subsequent figures indicates typical (median) error bars.
}
\end{figure}

\begin{figure}
\epsfig{file=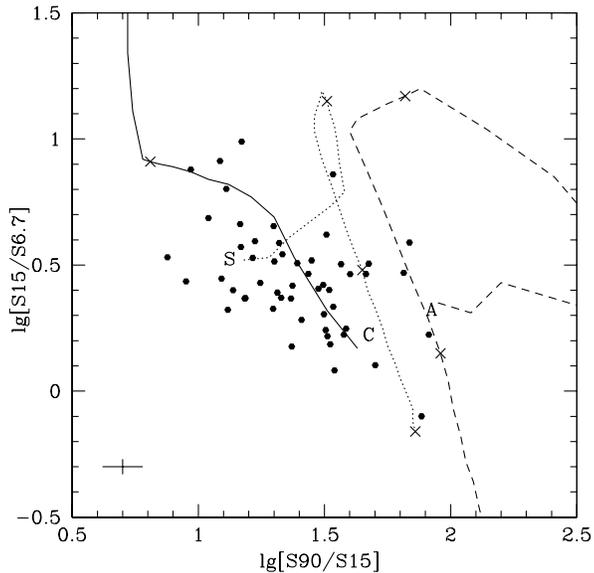,angle=0,width=8cm}
\caption{90-15-6.7 $\mu$m colour-colour diagram for sources in N2 and S1 detected in all 3 bands 
(all are galaxies), with loci for cirrus (C),
starburst (S) and Arp220 (A) seds, from $z =$ 0 to 3.
}
\end{figure}

\begin{figure}
\epsfig{file=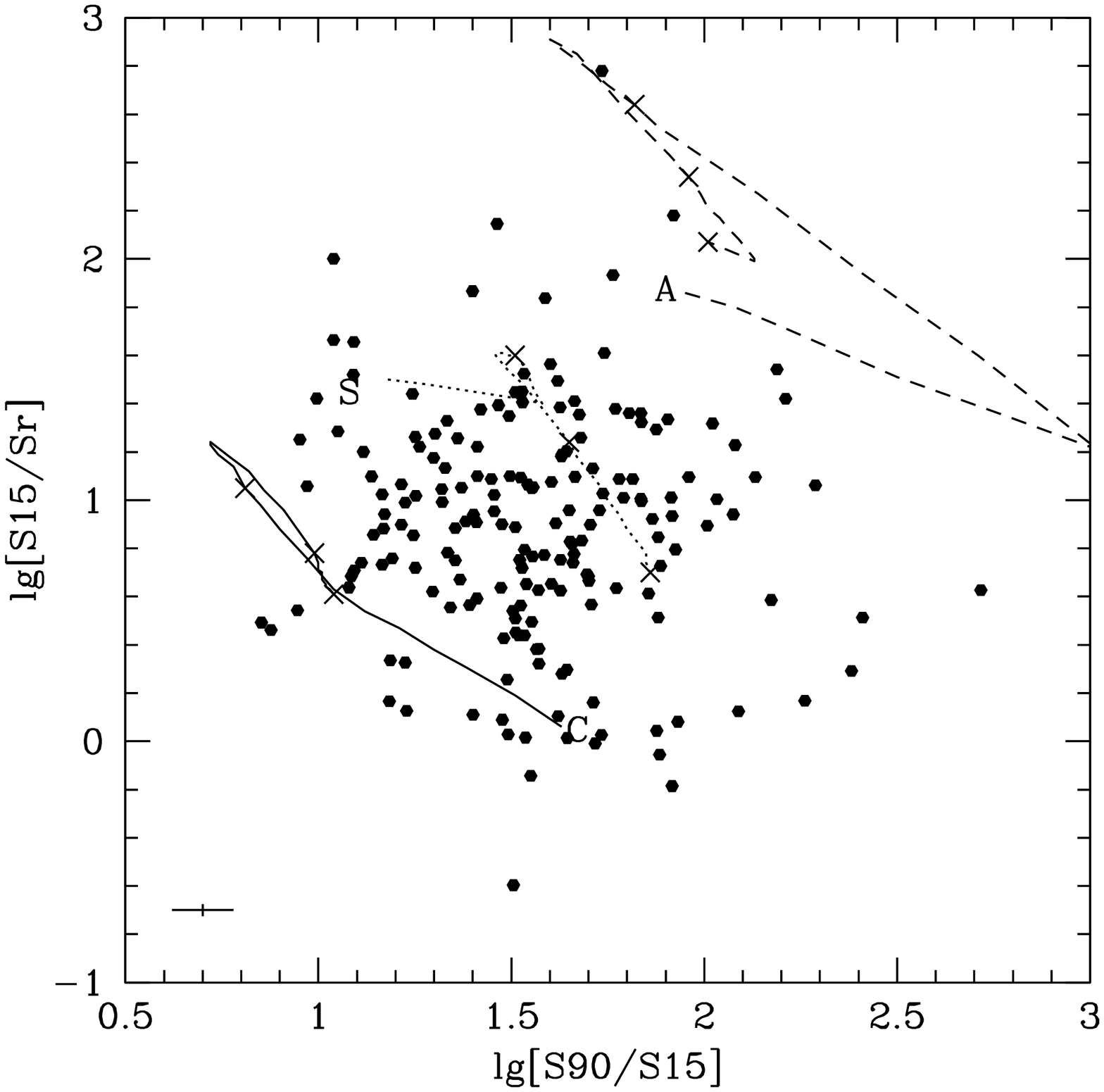,angle=0,width=8cm}
\caption{90-15-r $\mu$m colour-colour diagram for N1 , N2 and S1, with loci for cirrus (C),
starburst (S) and Arp220 (A) seds, from $z =$ 0 to 3.
}
\end{figure}

Optical colours provide a powerful discriminant between galaxies and
AGN.  Many of the Catalogue sources in N1 and N2 are detected in each
of the WFS band g', r', i'.  In figure 6 we show the $g'-r'-i'$
colour-colour diagram for sources classified by Sextractor as
galaxies.  The brighter galaxies define a very tight set of colours,
narrower in $(r'-i')$ than in (g'-r)', characteristic of galaxies with 0
$<$ z $<$ 0.5.  Model curves are shown for E, Sbc and starburst galaxies, for z = 0-2.
In figure 7 we show the corresponding plot for objects
classified as star-like (excluding actual stars), with different
symbols for sources which the photometric redshift code (see section
5) classifies as having AGN-type seds and those with galaxy seds.  The
AGN occupy quite a tight colour region centred on $(g-r)\sim$ 0.4,
$(r'-i') \sim$ 0.3.  Model curves are shown for E and starburst galaxies for z = 0-2 and for AGN
with z = 0-6.  Objects with galaxy seds show some overlap with Fig
6, but with more scatter to higher values of $(r'-i')$, consistent with
having higher redshifts.

Figure 8 shows $\lg(S_{15}/S_r)$ versus $(u-r)$ for ELAIS sources
, where $S_r$ denotes the r'-band flux in mJy.  
There is a clear separation between the Galactic
stars in the lower part of the diagram and the AGN and compact
galaxies in the upper part, and between the galaxies and AGN.   The objects classified as starlike but with galaxy seds have
values of $S_{15}/S_r$ up to 300 and these must be heavily obscured
starbursts like Arp 220, or Type 2 AGN.

\begin{figure}
\epsfig{file=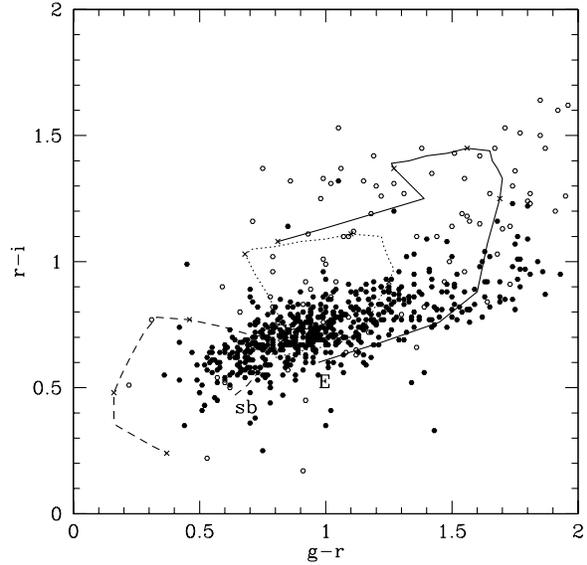,angle=0,width=8cm}
\caption{$g'-r'-i'$ colour-colour diagram for galaxies in N1 and N2.  
Filled circles: r' $<$ 21 mag., open circles: r' $>$ 21 mag.
Model loci for E (solid line), Sbc (dotted line) and starburst (broken line) galaxies are shown for z = 0-2
(crosses denote z = 0.5, 1, 1.5, 2).
}
\end{figure}

\begin{figure}
\epsfig{file=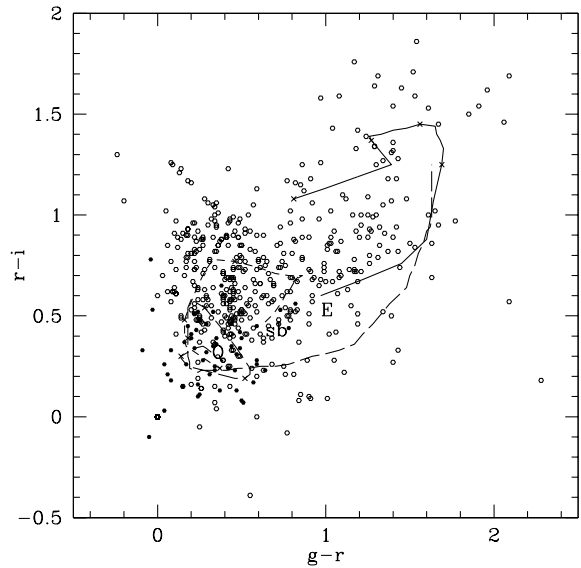,angle=0,width=8cm}
\caption{$g'-r'-i'$ colour-colour diagram for star-like objects in N1 and N2.  Filled circles: AGN sed, open
circles: galaxy sed.
Model loci for a quasar (long-dashed line, z = 0-6), E galaxy (solid line, z = 0-2), starburst (dashed line, z = 0-2),
are shown (crosses: z = 0.5, 1, 1.5, 2).
}
\end{figure}


\begin{figure}
\epsfig{file=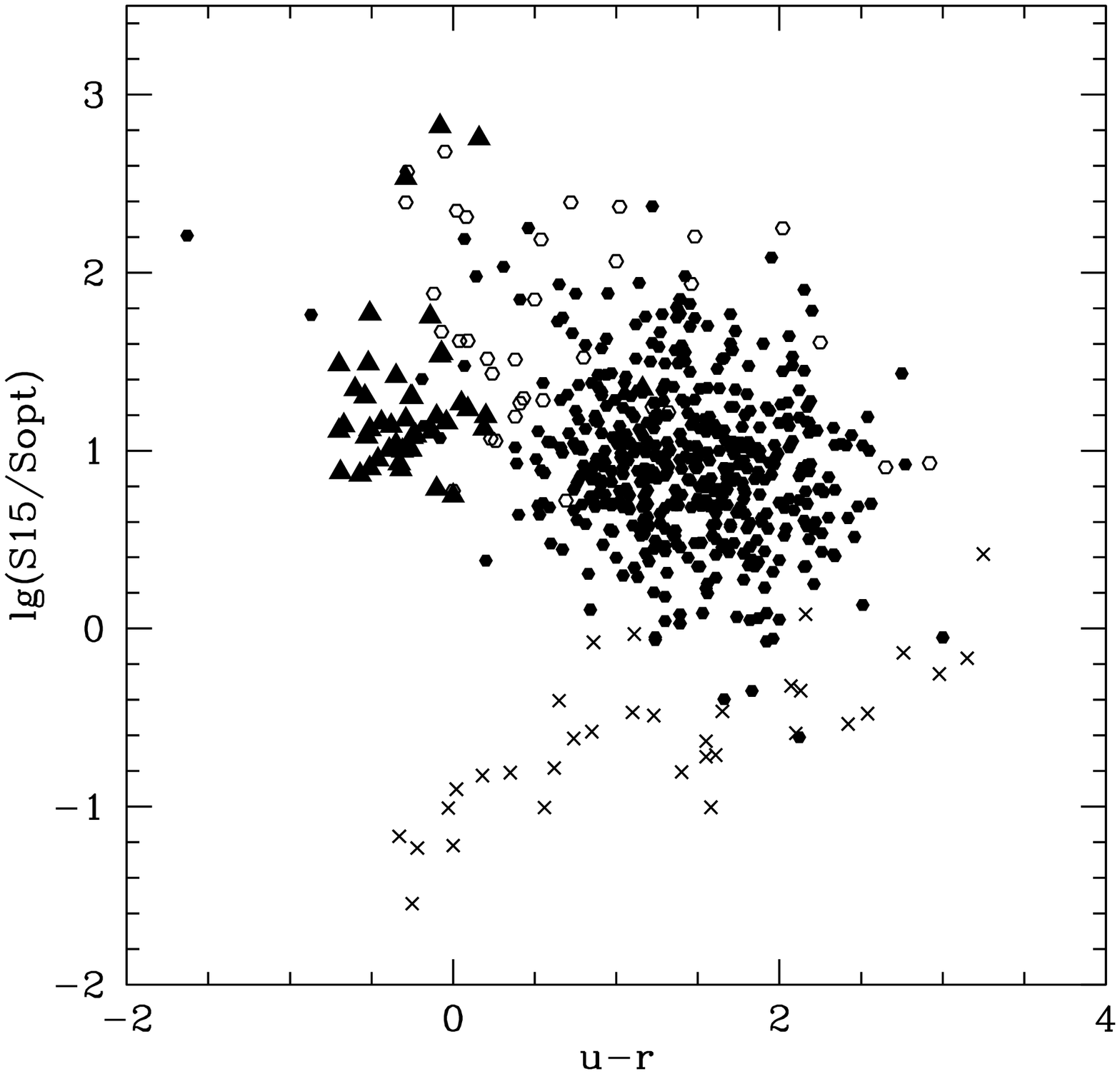,angle=0,width=8cm}
\caption{15/r versus $(U-r')$ for ELAIS galaxies and AGN in N1 and N2.
Filled circles: extended optical identifications, filled triangles:
star-like objects with AGN seds, open circles: star-like objects with
galaxy seds, crosses: Galactic stars.
Note that the (U-r') colours of the Galactic stars are inaccurate because of saturation effects.
}
\end{figure}

\section{Photometric redshifts in N1 and N2}

The WFS optical data in the U,g',r',i',Z bands, and J,H,K data (where available) for N1 and N2 allows us to determine photometric redshifts
for a large fraction of the sample.  The approach used is that of Rowan-Robinson (2003) with a set of
6 galaxy templates, and with the option of varying $A_V$. 
In addition two simple AGN templates are included,
based on the average optical QSO spectrum of Rowan-Robinson (1995),  
modified to take account of observed seds of ELAIS AGN
(details of all the templates used are given at http://astro.ic.ac.uk/$\sim$mrr/photz).
Figs 9, 10 show the comparison of the data for AGN with spectroscopic redshifts with the assumed 
templates.
The application of this code to the full
WFS data set is described by Babbedge et al (2004).   We have also used the UBRI data in S2 (Pozzi et al 2003)
to estimate photometric redshifts.  In S1 we have only BRI data in the optical and the photometric calibration at B is
uncertain.  We found that without U (or u) data, it was impossible to determine photometric redshifts
for quasars, but results for galaxies were still good, and these are included in the Catalogue provided at least 3
photometric bands between B and K are available.
Figure 11 shows the comparison
of photometric and spectroscopic redshifts for galaxies in N1, N2, S1 and S2.  The spectroscopic redshifts in N1 and N2 
are reported by
Perez-Fournon et al (2004) and Serjeant et al (2004), in S1 by La Franca et al (2004) and in S2 by Pozzi et al (2003).  
The agremeent is good, within the uncertainty of the photometric method ($\sim 10 \%$ in (1+z) according to the
analysis of Rowan-Robinson (2003)).  A fuller discussion of the application of photometric redshift techniques to
the WFS data is given by Babbedge et al (2004).
Of course, it needs to be emphasized, that a 10 $\%$ accuracy in (1+z) means that photometric redshifts $<<$ 1 will be very inaccurate.
One of the two strongly discrepant points in Fig 11 is a case where the photometric redshift is determined from only 3 photometric
bands, so as in Rowan-Robinson (2003) we can claim excellent performance for the photometric redshift method
if 4 bands are available.  Figure 12 shows the corresponding plot for AGN in N1, N2 and S2: the results are surprisingly good.
Several AGN acquire spurious photometric redshifts in the range 2-2.5 because their U magnitudes are fainter than
predicted by the template, possibly because of the effect of extinction, so they are interpreted as Lyman drop-outs.
The code is quite successful both in finding the AGN and estimating their redshifts.
Out of 33 sources with both spectroscopic and photometric redshifts, which are classified either spectroscopically
or photometrically as AGN, only 2 are not classified spectroscopically as AGN, presumably due to aliassing in the
photometric solution.  5 spectroscopic AGN are not picked up as AGN photometrically though all have
consistent photometric redshift estimates, perhaps due a weak AGN continuum.
Examples of the template fits to optical and near ir data can be seen
in Figs 26-28.

\begin{figure}
\epsfig{file=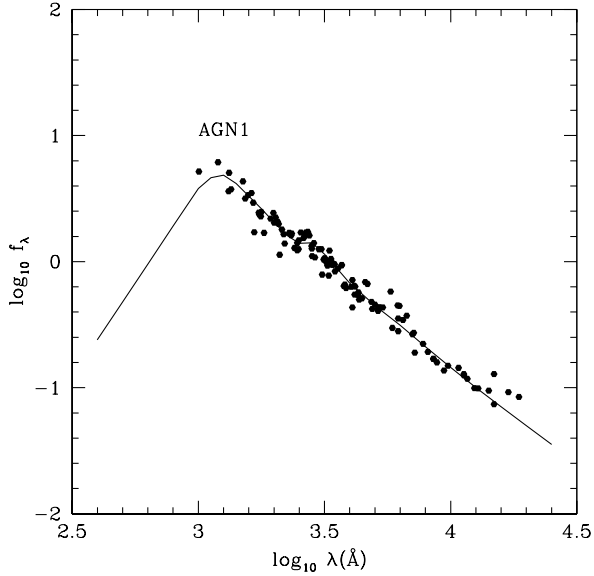,angle=0,width=8cm}
\caption{Comparison of photometric data for AGNs with $n_{typ}$ = 7 and known spectroscopic
redshifts with the assumed template. 
}
\end{figure}

\begin{figure}
\epsfig{file=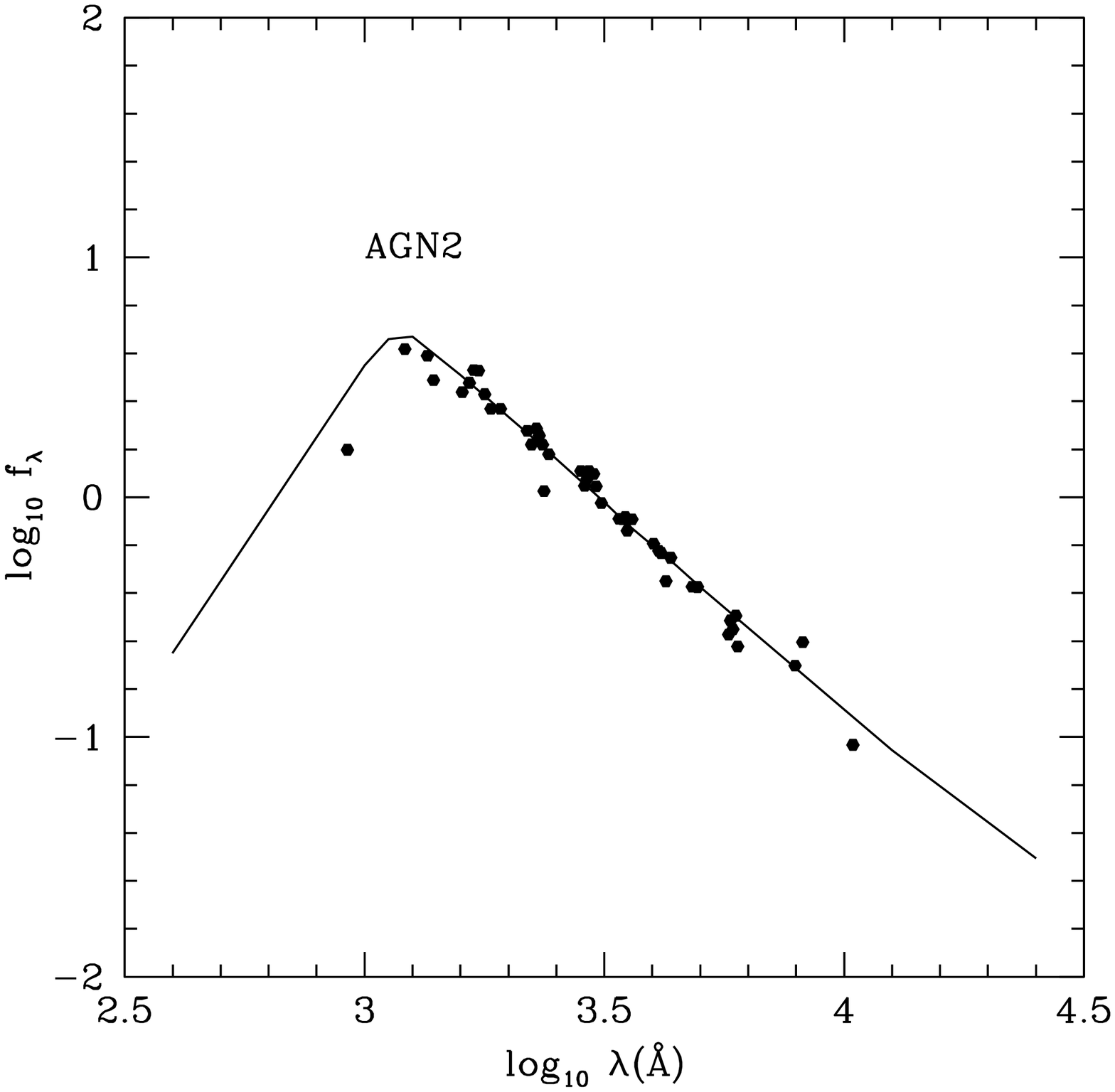,angle=0,width=8cm}
\caption{Comparison of photometric data for AGNs with $n_{typ}$ = 8 and known spectroscopic
redshifts with the assumed template.  
}
\end{figure}

\begin{figure}
\epsfig{file=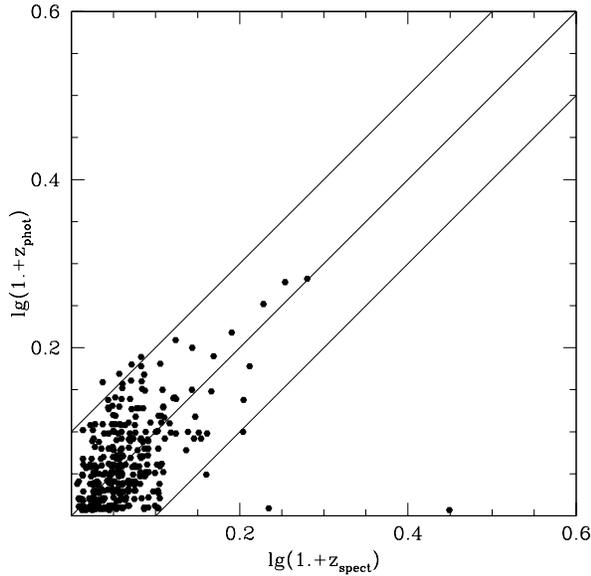,angle=0,width=8cm}
\caption{$\lg_{10}(1+z_{phot})$ versus  $\lg_{10}(1+z_{spect})$ for galaxies. 
The straight lines bracket the range $\Delta lg_{10}(1+z) = \pm 0.1$ ($\pm 2.5\sigma$ according to the analysis of 
Rowan-Robinson (2003)).  The photometric redshifts are
The straight lines indicate the $\pm 2.5 \sigma$ range.  The photometric redshifts are
for an assumed $A_V$ = 0.
}
\end{figure}

\begin{figure}
\epsfig{file=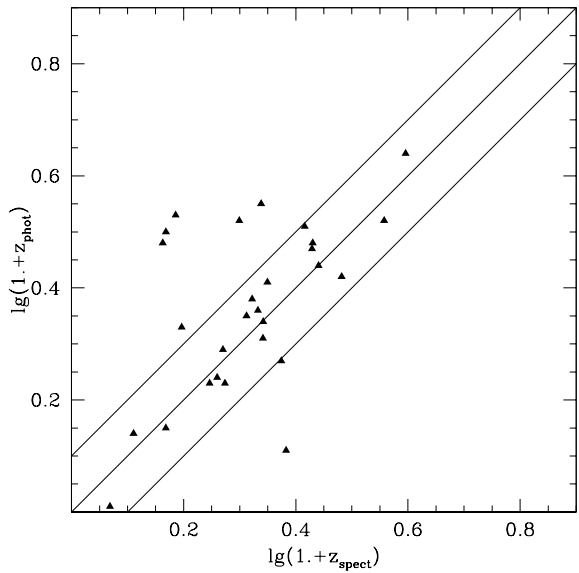,angle=0,width=8cm}
\caption{$\lg_{10}(1+z_{phot})$ versus  $\lg_{10}(1+z_{spect})$ for AGN. 
}
\end{figure}

Figures 13-16 show redshift histograms at 15, 90 and 175 $\mu$m and 20 cm, for sources brighter than the characteristic depth
specified in Table 1, with the dotted lines in Figs 11 and 14
indicating spectroscopic redshifts.  The broken histogram indicates the effect of assigning the 15 $\mu$m blank-field sources
in S1 (which have R $>$ 20) redshifts uniformly distributed in the range 0.2-1.5, and the blank-field sources in N1 and N2
(which have r' $>$ 24) redshifts uniformly distributed in the range 0.6-1.5, based on the R-z diagram of LaFranca et al
(2003, their Fig 15).  The median redshift is 0.30 at 20 cm, 0.17 at 15 $\mu$m and 0.10 at 
90 and 175 $\mu$m.
In their analysis of the 15 $\mu$m redshift distribution in S1, La Franca et al (2003) and Pozzi et al (2004)
infer a stronger secondary peak at around z = 1 than is indicated here.  Model predictions by Rowan-Robinson (2001)
at 15, 90 and 175 $\mu$m are shown in Figs 13-15, and by Pozzi et al (2003) at 15 $\mu$m in Fig 13.  The overall
agreement with the Rowan-Robinson (2001) model is reasonable, though the Pozzi et al (2003b) model may be a slightly
better fit at 15 $\mu$m.  This merits further investigation.

The luminosity function and evolution at 15 $\mu$m are discussed by La Franca et al (2004), Pozzi et al (2004),
Perez-Fournon et al (2004), at 90 $\mu$m by Serjeant et al (2004) and at 175 $\mu$m by Perez-Fournon et al (2004).

\begin{figure}
\epsfig{file=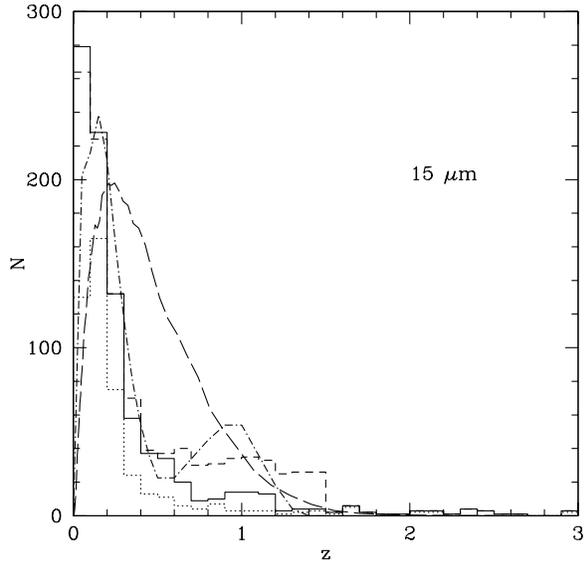,angle=0,width=8cm}
\caption{Redshift histogram for ELAIS 15 $\mu$m sources.  Solid curve: both photometric and spectroscopic redshifts (840 sources),
dotted curve: spectroscopic redshifts only (468 sources), broken curve: effect of assigning blank fields uniformly to range $0.5 < z < 1.5$.
Long broken curve: predicted redshift distribution from Rowan-Robinson (2001), dash-dotted curve: predicted redshift
distribution from Pozzi et al (2004).
}
\end{figure}

\begin{figure}
\epsfig{file=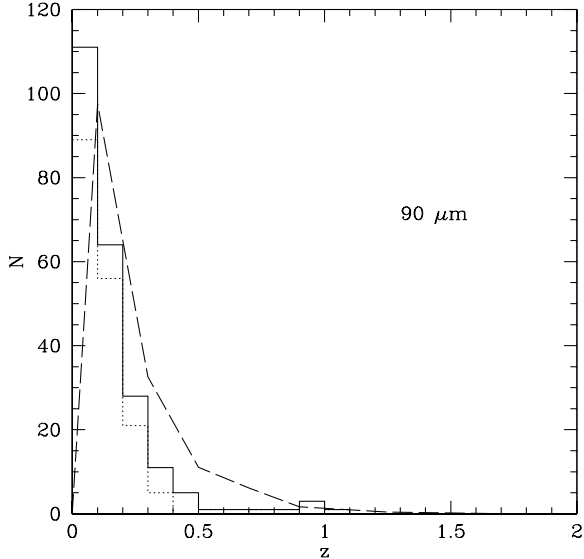,angle=0,width=8cm}
\caption{Redshift histogram for ELAIS 90 $\mu$m sources.  Solid curve: both photometric and spectroscopic redshifts (229 sources),
dotted curve, spectroscopic redshifts only (175 sources),
Long broken curve: predicted redshift distribution from Rowan-Robinson (2001).
}
\end{figure}

\begin{figure}
\epsfig{file=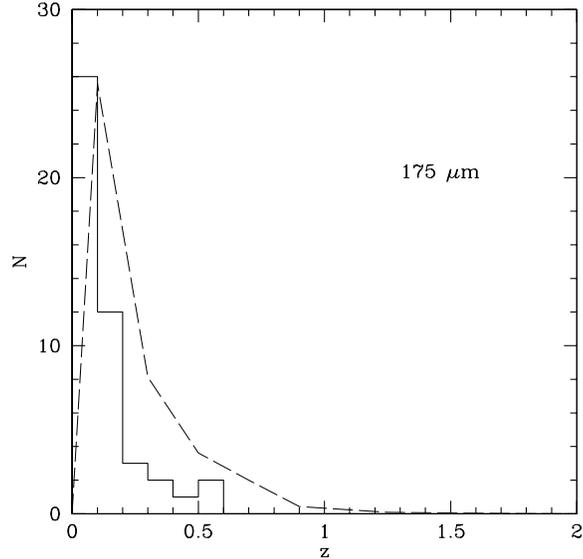,angle=0,width=8cm}
\caption{Redshift histogram for ELAIS 175 $\mu$m sources.  Solid curve: both photometric and spectroscopic redshifts (46 sources).
Long broken curve: predicted redshift distribution from Rowan-Robinson (2001).
}
\end{figure}

\begin{figure}
\epsfig{file=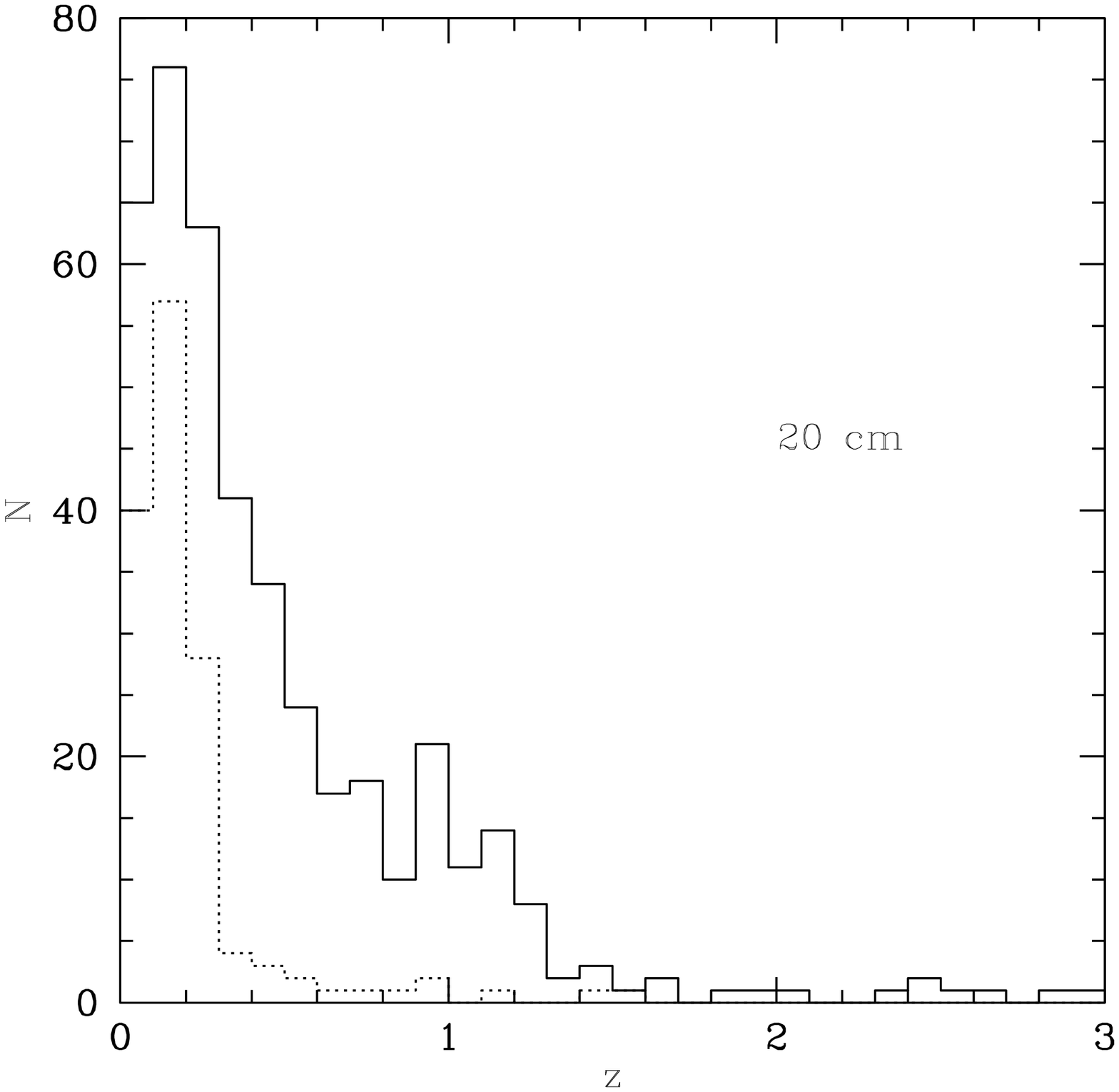,angle=0,width=8cm}
\caption{Redshift histogram for ELAIS radio sources.  Solid curve: both photometric and spectroscopic redshifts (424 sources),
dotted curve: spectroscopic redshifts only (140 sources).
}
\end{figure}

Figure 17 shows $\lg(S_{15}/S_r)$ versus z, where $z = z_{best}$, for objects classified by Sextractor as galaxies, 
together with
predicted loci for cirrus, M82 and Arp220 starbursts.  Although most of the galaxies have  $z<$ 0.3, there are
an interesting subset with 0.6 $<z<$ 1.2.  
Figure 18 shows the corresponding diagram for star-like objects,
with model loci for AGN dust tori, and M82 and Arp220 starbursts. The objects with AGN seds, shown as filled
circles, follow the (Type 1) AGN dust torus model line well.  Most
with galaxy seds (and some from Fig 17 with  0.2 $<$ z $<$ 0.8)
have $S_{15}/S_r$ values higher than any of the model loci.  These may represent a new population of heavily obscured starbursts
or Type 2 AGN.

Fig 19 shows $\lg(S_{15}/S_{20 cm})$ versus z for all sources, for an assumed rest-frame
$S_{60}/S_{20cm}$ ratio of 200.  There is a dependence of $S_{15}/S_{20 cm}$ on infrared sed type and on redshift.
Thus even if there is a perfect correlation between 60 $\mu$m and 20 cm flux, we do not expect this correlation to be
preserved at other infrared wavelengths.  Note the two radio-loud AGN in the lower part of Fig 19.
The 15 $\mu$m - radio correlation has also been discussed by Gruppioni et al (2003).

Fig 20 shows $\lg(S_{15}/S_{6.7})$ versus redshift for all sources in N2 and S1.
Values of $S_{15}/S_{6.7}$ $<$ 1 are predominantly stars and have been correctly recognized as such by the photometric
redshift code.  The 15/6.7 colour ratio is not a good discriminant between the galaxy models at low redshift, though it is
clear that at high redshift the sources conform well to the AGN dust torus locus.

Figures 21 and 22 show $\lg(S_{90}/S_r)$ and $\lg(S_{175}/S_r)$ versus z, with model loci for cirrus, M82 
and Arp220 starbursts.
The models nicely bracket the observational points.  However a plot of $\lg(S_{175}/S_{90})$ versus $\lg(1+z)$ (Fig 21)
shows that at low redshifts there is a group of nearby galaxies with colder dust ( $\nu B_{\nu}(T)$ colour temperatures
in the range 18-25 K) than the cirrus template used
here (which matches IRAS colours for cirrus sources well).  Clearly we do not expect all the dust in a galaxy
to be at a single temperature.  As we move to longer wavelengths we expect to see emission from cooler dust
further from the centre of the galaxy.  Observations of these galaxies at 350-850 $\mu$m would be valuable
for understanding the true distribution of dust mass and temperature in galaxies.

\begin{figure}
\epsfig{file=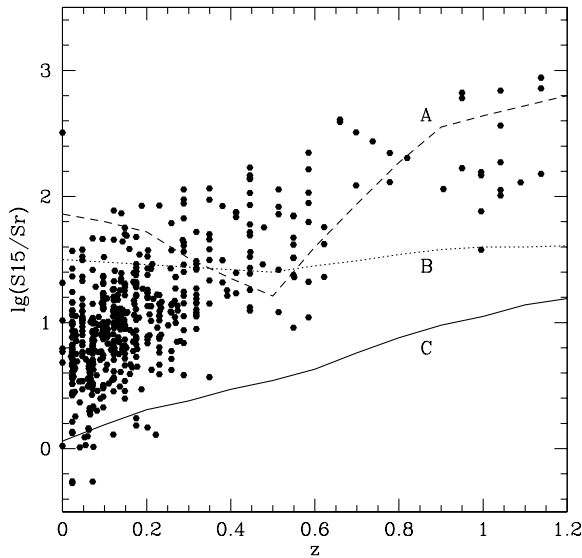,angle=0,width=8cm}
\caption{15$\mu$m/$r$-band colour versus z for ELAIS galaxies , with loci for cirrus (C, solid line),
starburst (B, dotted line) and Arp220 (A, broken line) seds.
}
\end{figure}

\begin{figure}
\epsfig{file=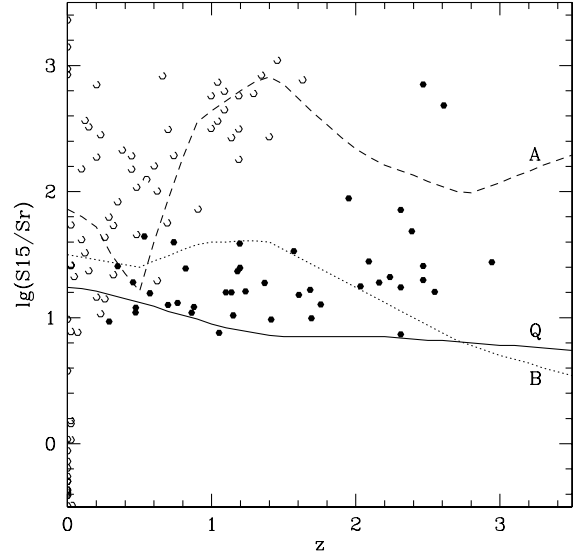,angle=0,width=8cm}
\caption{ 15$\mu$m/$r$-band colour versus z for ELAIS star-like sources (excluding stars), with loci for 
AGN (Q, solid line),
starburst (B, dotted line) and Arp220 (A, broken line) seds.  Filled circles: sources with AGN optical seds, 
open circles: sources the galaxy seds.
}
\end{figure}

\begin{figure}
\epsfig{file=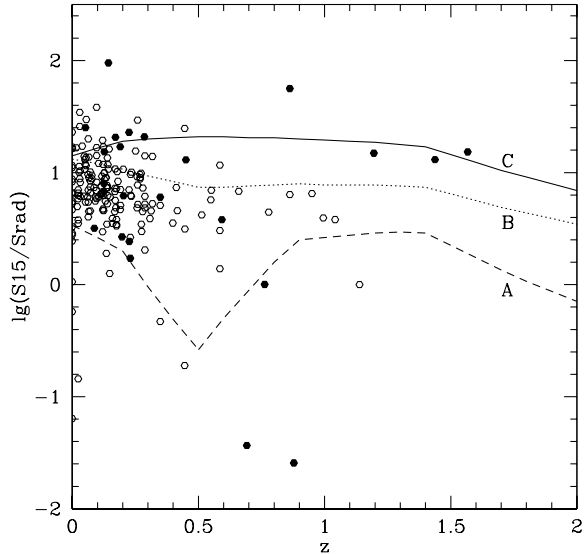,angle=0,width=8cm}
\caption{ 15$\mu$m/20cm colour versus z for ELAIS Catalogue sources, with loci for cirrus (C, solid line),
starburst (B, dotted line) and Arp220 (A, broken line) seds.  Filled circles: AGN, open circles: galaxies.
}
\end{figure}
\begin{figure}
\epsfig{file=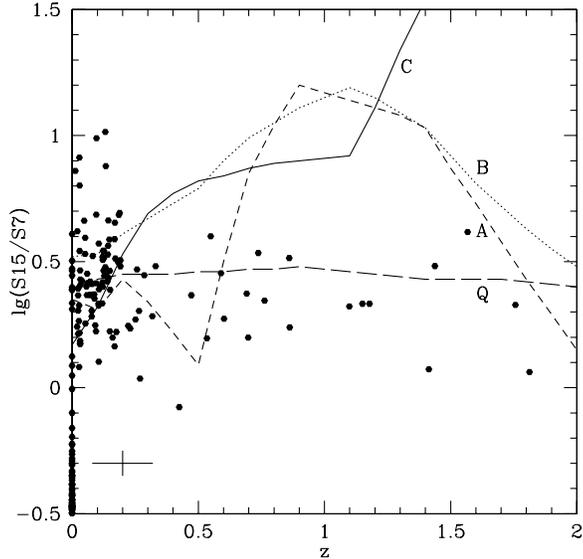,angle=0,width=8cm}
\caption{15/6.7 $\mu$m, colour versus z for N2 and S1, with loci for cirrus (C, solid line),
starburst (B, dotted line), Arp220 (A, short broken line) and AGN dust tori (Q, long broken line) seds.
}
\end{figure}

\begin{figure}
\epsfig{file=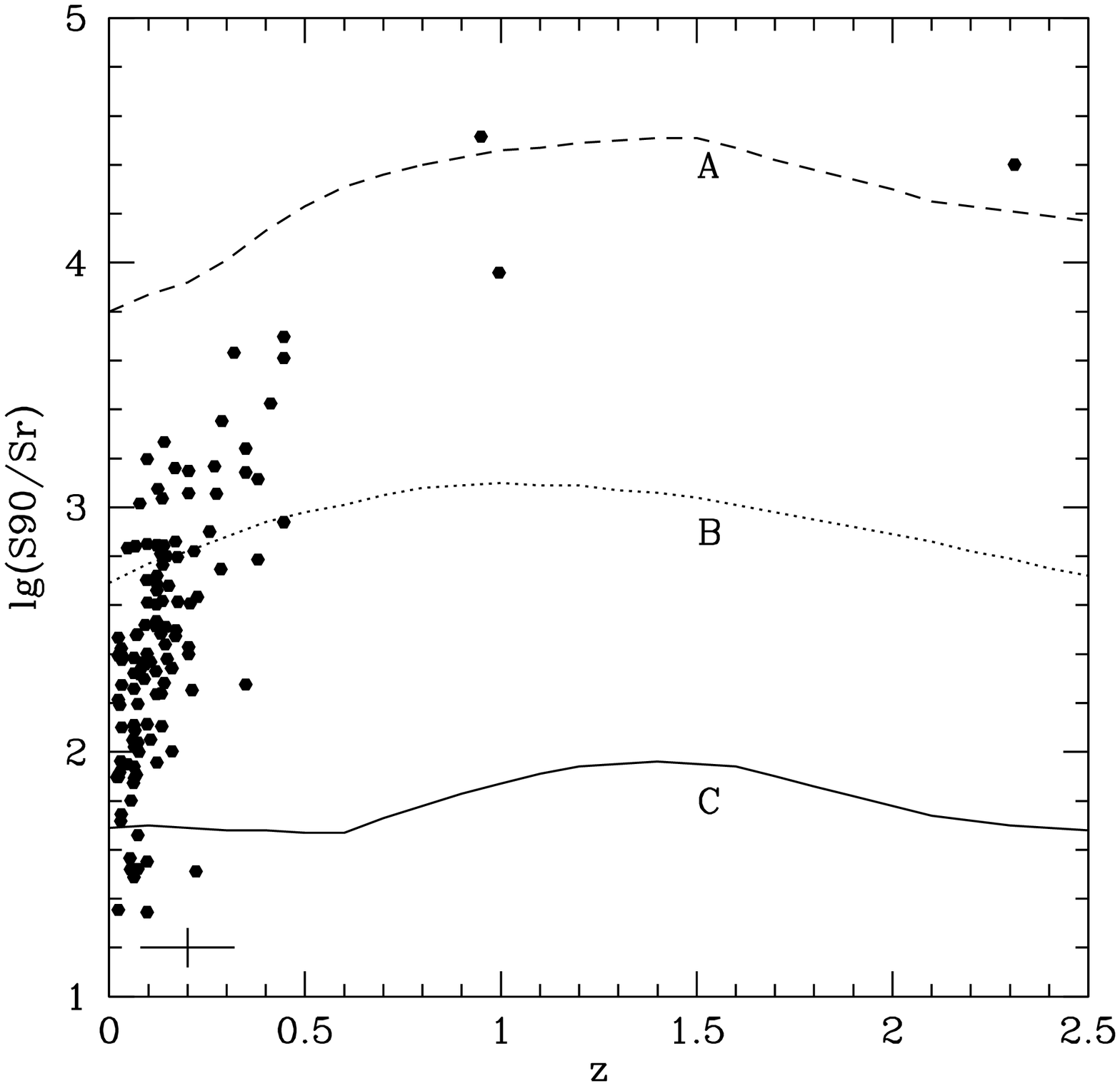,angle=0,width=8cm}
\caption{90 $\mu$m/$r$-band colour versus z for ELAIS galaxies, with loci for cirrus (C, solid line),
starburst (B, dotted line) and Arp220 (A, broken line) seds.
}
\end{figure}

\begin{figure}
\epsfig{file=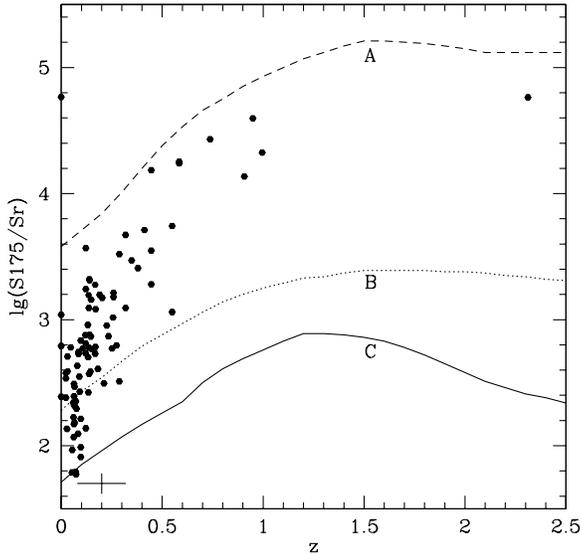,angle=0,width=8cm}
\caption{175$\mu$m/$r$-band colour versus z for galaxies in N1 and N2, with loci for cirrus (C, solid line),
starburst (B, dotted line) and Arp220 (A, broken line) seds.
}
\end{figure}

\begin{figure}
\epsfig{file=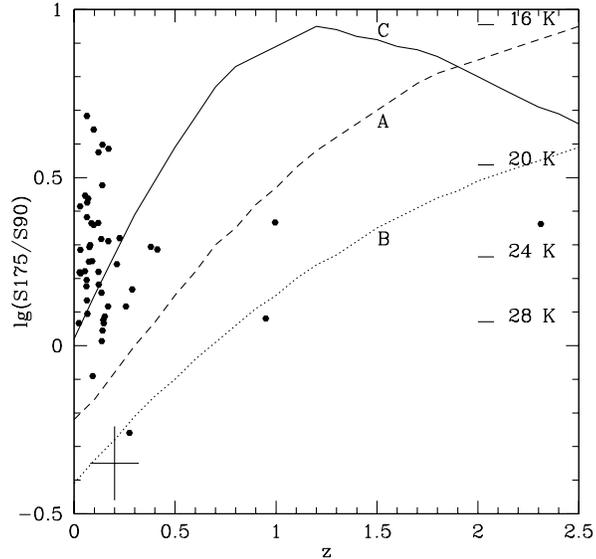,angle=0,width=8cm}
\caption{175/90 $\mu$m, colour versus z for galaxies in N1 and N2, with loci for cirrus (C, solid line),
starburst (B, dotted line) and Arp220 (A, broken line) seds.
Horizontal bars at right indicate dust colour temperatures from a $\nu B_{\nu}(T)$ model.
}
\end{figure}

\begin{figure}
\epsfig{file=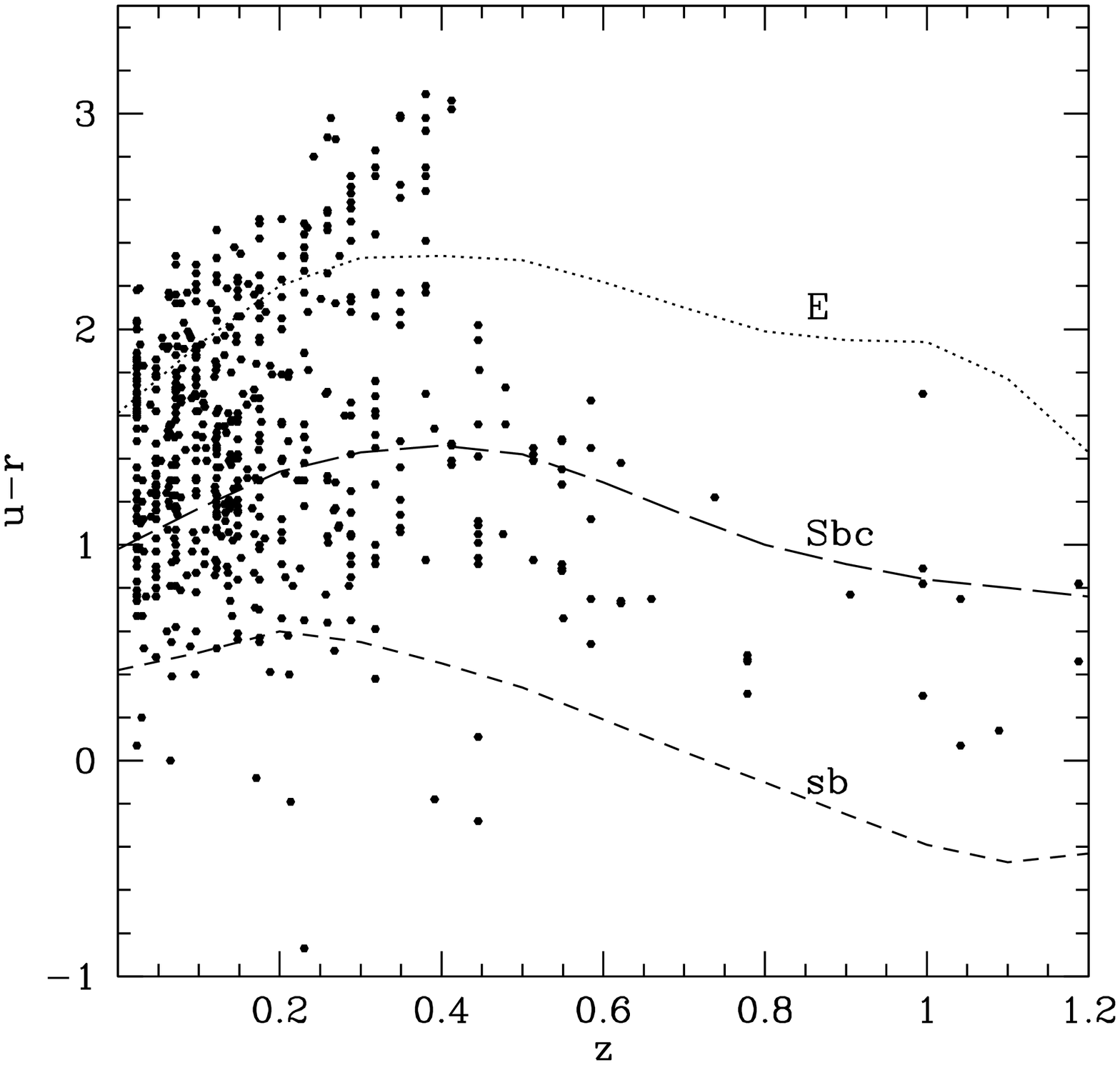,angle=0,width=8cm}
\caption{$U-r'$ versus versus z for galaxies in N1 and N2.  
Model loci are for an E galaxy (dotted line), Sbc galaxy (long-dashed line) and starburst (dashed line).
}
\end{figure}

Figure 24 shows $(u-r)$ versus $\lg_{10}(1+z)$ for objects classified as extended, compared with model predictions
for E, Sbc and starburst galaxy seds.  The photometric redshift code suggests that galaxies with (u-r) $>$ 2 
show evidence of extinction, with $A_V$ in the range 1-3.  
Figure 25 shows the corresponding diagram for star-like objects.
The objects with AGN-type seds follow a narrow locus, with objects with spectroscopic redshifts (filled circles)
agreeing well with the adopted AGN sed,
while those with galaxy seds
occupy a similar region to the galaxies of Fig 24.

A full discussion of the JHK, 6.7 and 15 $\mu$m colour-colour diagrams has been given by Vaisanen et al (2002)
and the results are not significantly altered by the larger and more accurately calibrated sample here.
The JHK data help the photometric redshift fitting but do not help understand the nature of the dust-emission
components in galaxies.

\begin{figure}
\epsfig{file=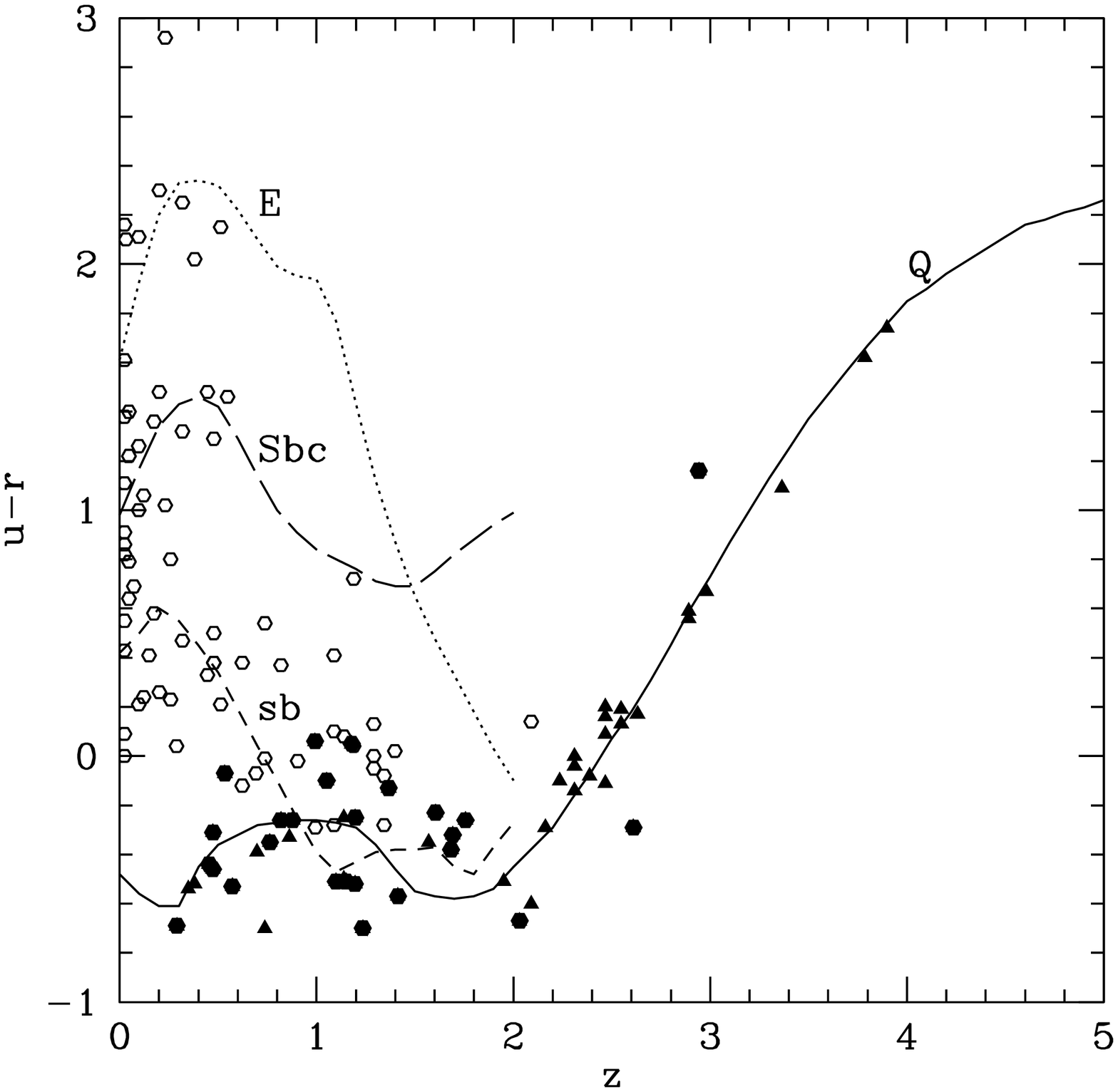,angle=0,width=8cm}
\caption{$U-r'$ colour versus z for star-like objects in N1 and N2.  Filled triangles denote objects with AGN seds, 
Filled circles are spectroscopic AGN (Sy 1 and 2) and
open circles are objects with galaxy seds.
The model loci are for an AGN (solid line), E galaxy (dotted line), Sbc galaxy (long-dashed line) and starburst
(dashed line).
}
\end{figure}

\section{Infrared spectral energy distributions}

For sources detected in multiple ISO bands we can compare the infrared spectral energy distributions
with those for standard model templates.  Figure 26 compares the rest-frame seds of 9 galaxies detected
in all four ISO bands with a standard cirrus emission spectrum (Efstathiou and Rowan-Robinson
2003: surface brightness parameter $\psi$ = 5, age of starburst = 5 Gyrs).  All have modest 
infrared luminosities ($L_{ir} < 11.5$).  Figure 27 shows a similar plot for luminous cirrus
galaxies ($L_{ir} > 11.5$).  These represent an interesting new population of luminous cool galaxies,
with redshifts in the range 0.15-0.5, consistent with the strong evolution postulated for
cirrus galaxies by Rowan-Robinson (2001) and
perhaps related to the very luminous cirrus galaxies postulated by
Efstathiou and Rowan-Robinson (2003) to explain some of the high redshift
SCUBA galaxies.  For 160552.5+540650 there is evidence also for a
starburst component.
Figure 28 compares 2 galaxies detected in 4 ISO bands and 4 detected in 3 ISO bands with
a high optical depth starburst model which gives a good fit to Arp 220
(Efstathiou et al 2000: $A_V$ = 200, age of starburst = 26 Myr).
The tend to have higher infrared luminosities and optical sed type $n_{typ}$
= 1 or 2 (E or Sab),
consistent with the optical sed of Arp220, which shows very little contribution from young massive stars
due to the high extinction in this galaxy.
Figure 29 compares 3 galaxies detected in 3 ISO bands with an M82-type starburst
model (Efstathiou and Rowan-Robinson 2001: $A_V$ = 50, age of starburst = 26 Myr).
Two galaxies for which an AGN dust torus component is also required are included,
as well as a further two objects which only require an AGN dust torus component.
These figures include all ultraluminous ELAIS galaxies detected in 3 bands ( $L_{ir} > 12.22$ ).
The four infrared sed model components used in Figs 26-29 are the same as
those used in count models by Rowan-Robinson (2001).

\begin{figure}
\epsfig{file=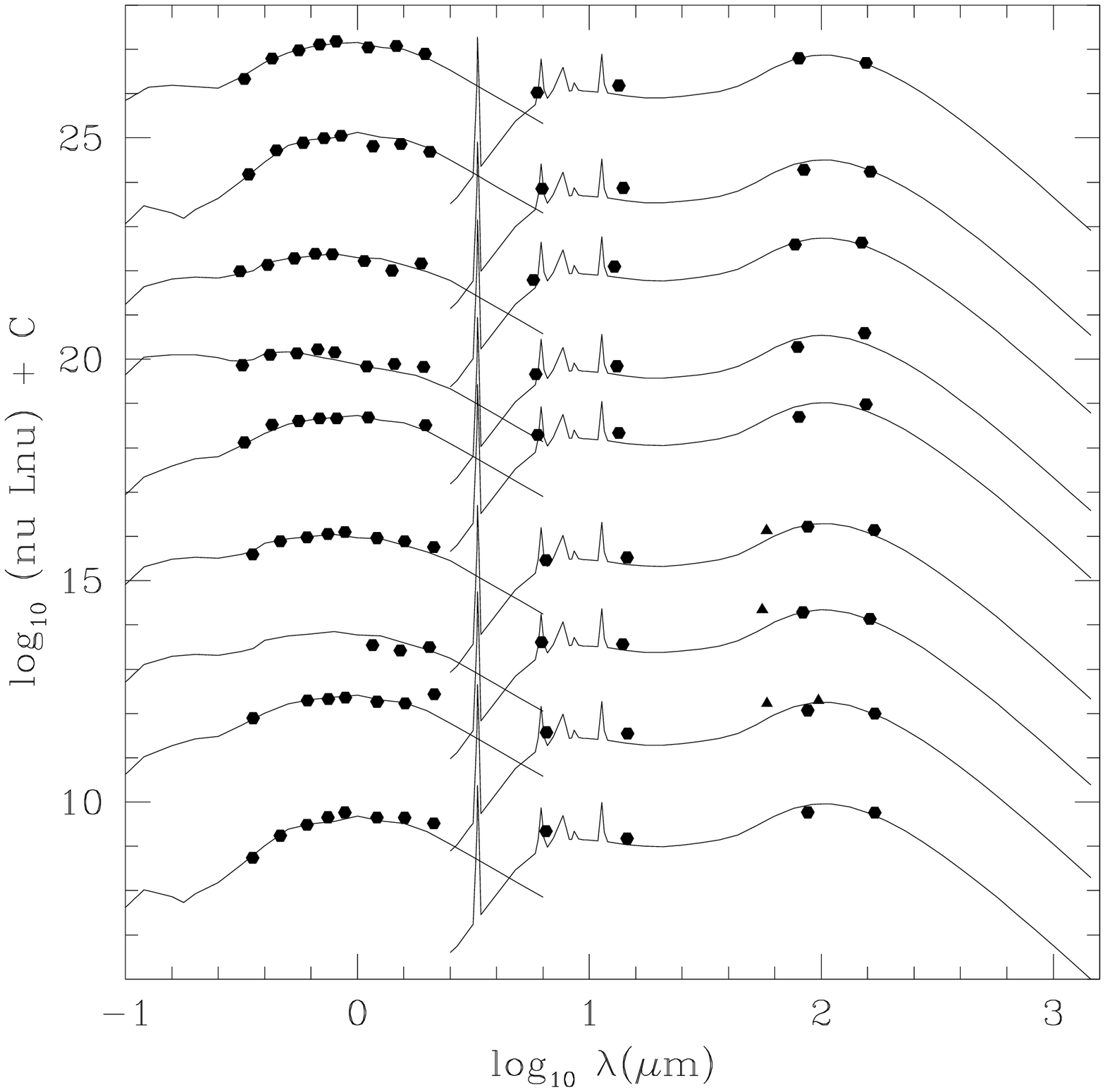,angle=0,width=8cm}
\caption{9 ELAIS galaxies detected in all 4 ISO bands for which a standard cirrus
model provides an excellent fit to the spectral energy distribution.
Filled circles: ISO data, filled triangles: IRAS data.
From bottom: ELAISC15 163359.2+405303 ($L_{ir}$=10.13, $n_{typ}$=1), 163401.8+412052 (10.42, 3), 
163506.1+411038 (11.01, (4)), 163525.1+405542 (10.46, 4), 163546.9+403903 (11.19, 3),
163548.0+412819 (11.21, 5), 163607.7+405546 (11.41, 4), 163613.6+404230 (10.67, 1), 163641.1+413131 (11.04, 2).
}
\end{figure}

\begin{figure}
\epsfig{file=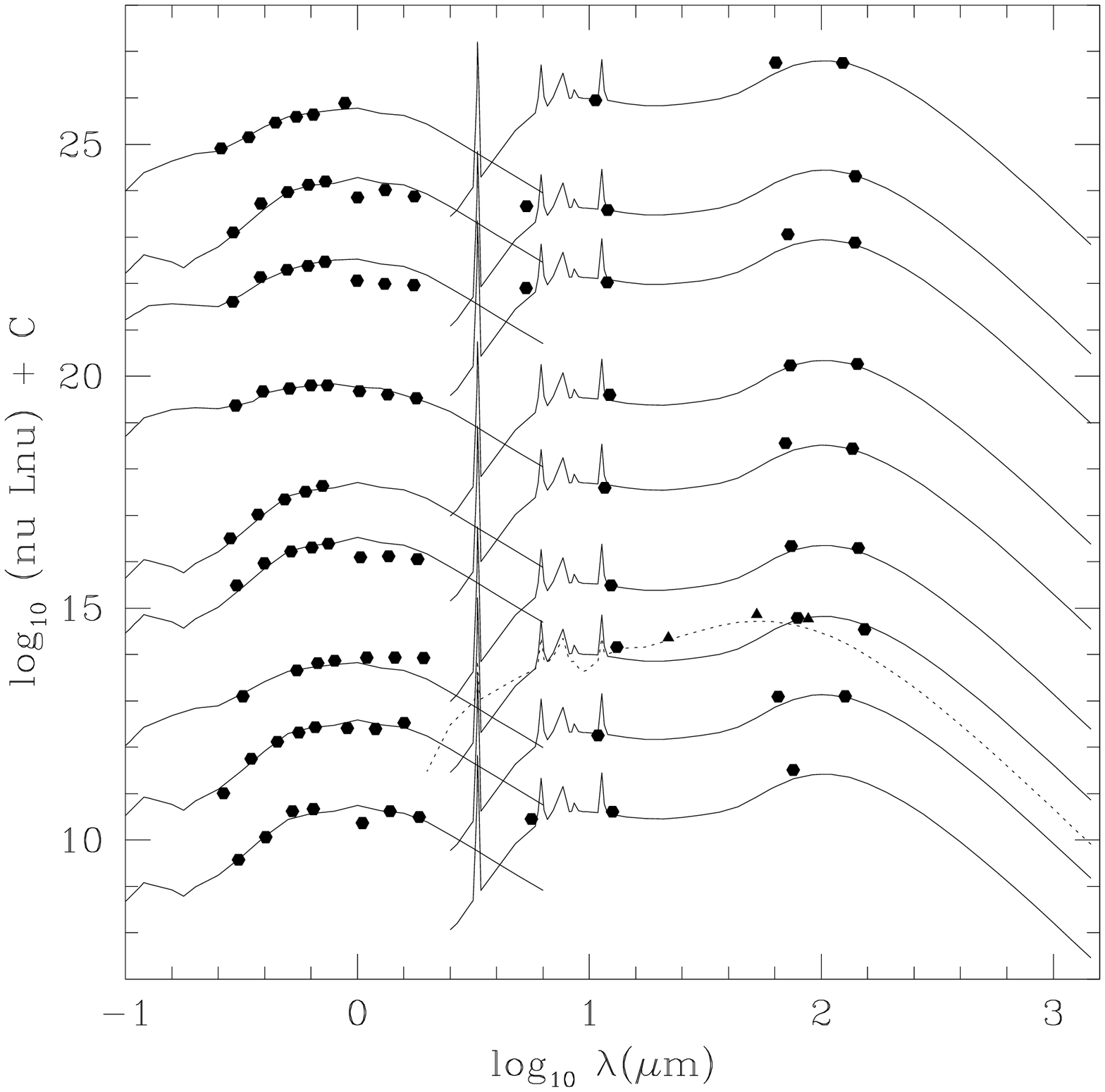,angle=0,width=8cm}
\caption{9 luminous ELAIS galaxies detected in 3 ISO bands for which a standard cirrus
model provides an excellent fit to the spectral energy distribution.
Filled circles: ISO data, filled triangles: IRAS data.
From bottom: ELAISC15 050225-3041112 ($L_{ir}$=11.59, $n_{typ}$=1), 160443.5+543332 (12.30, 1), 
160552.5+540650 (11.99, 3), 160553.3+542225 (11.52, 1), 160945.7+534944 (11.68, 1), 161041.2+541032 (11.51, 4),
163242.4+410847 (12.11, 2), 163449.5+412048 (11.61, 1), 163741.3+4111913 (11.97, 3).
An additional starburst component is shown for 160552.5+540650 (dotted curve).
}
\end{figure}

\begin{figure}
\epsfig{file=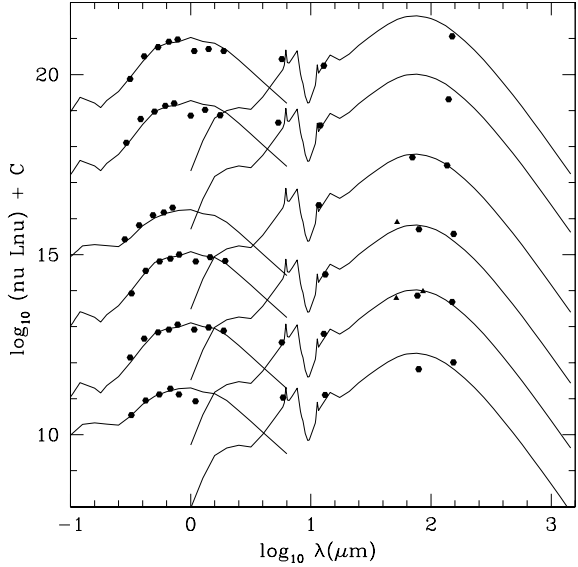,angle=0,width=8cm}
\caption{6 ELAIS galaxies detected in 3 or 4 ISO bands for which an Arp200
model provides an excellent fit to the spectral energy distribution.
From bottom: 163412.0+405652 (11.40, 2), 163608.1+410507 (12.16, 1), 
160934.7+53220 (11.46, 1), 161300.8+544838 (11.98, 2) , 
163449.5+412048 (12.15, 1), 163708.1+412856 (11.77, 1).
}
\end{figure}

\begin{figure}
\epsfig{file=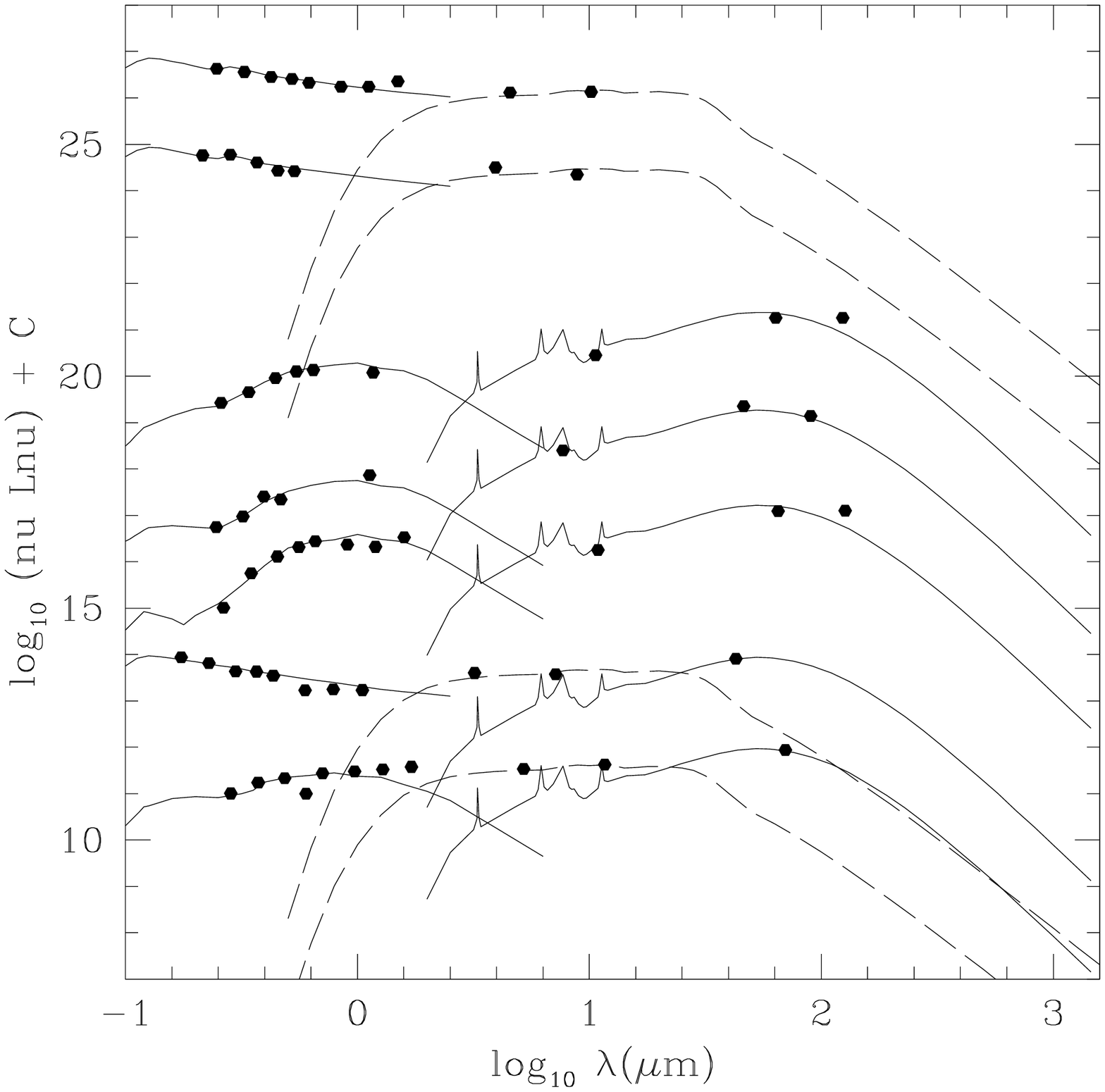,angle=0,width=8cm}
\caption{5 ELAIS galaxies detected in 3 ISO bands compared with an M82 starburst
model.  For two of the galaxies an AGN dust torus component has been included
to fit the mid-ir data (model from Rowan-Robinson 1995).
The top two loci are sources identified with quasars, with evidence only for dust torus emission
in the infrared.
From bottom: ELAIS 163751.4+413027 (12.20, 4) , 164010.1+410521 (13.17, 8), 160443.5+543332 (12.41, 1), 
163615.7+404759 (13.00, 2), 163741.3+411913 (12.11, 3),
163352.4+402112 (11.79, 7), 163502.7+412953 (11.48, 7).
}
\end{figure}

For all ELAIS galaxies detected in two or more bands and with spectroscopic or photometric redshifts
(a total of 306 galaxies) we have selected the best-fitting of the four model components cirrus,
M82, Arp220 and AGN dust torus, and estimated the bolometric infrared luminosity
(from 3-1000 $\mu$m: here L denotes the $log_{10}$ of the luminosity).  
We have used the 15 $\mu$m flux (or upper limit) to estimate the luminosity in an
AGN dust torus component, $L_{tor}$, if the 15 $\mu$m emission is interpreted as due to
such a component (without corroboration at longer wavelengths, $L_{tor}$ gives
a more conservative estimate of luminosity than starburst or cirrus models).
We estimated the corresponding optical bolometric luminosities (0.04-3 $\mu$m), using the photometric 
sed templates
(Rowan-Robinson 2003).  Figure 30 shows a plot of ($L_{ir} - L_{opt}$) versus $L_{ir}$. For ease of
comparison with earlier work, an Einstein de Sitter model ($\Omega$ = 1, $\Lambda$ = 0) model,
with $H_0$ = 50, is used. 

\begin{figure*}
\epsfig{file=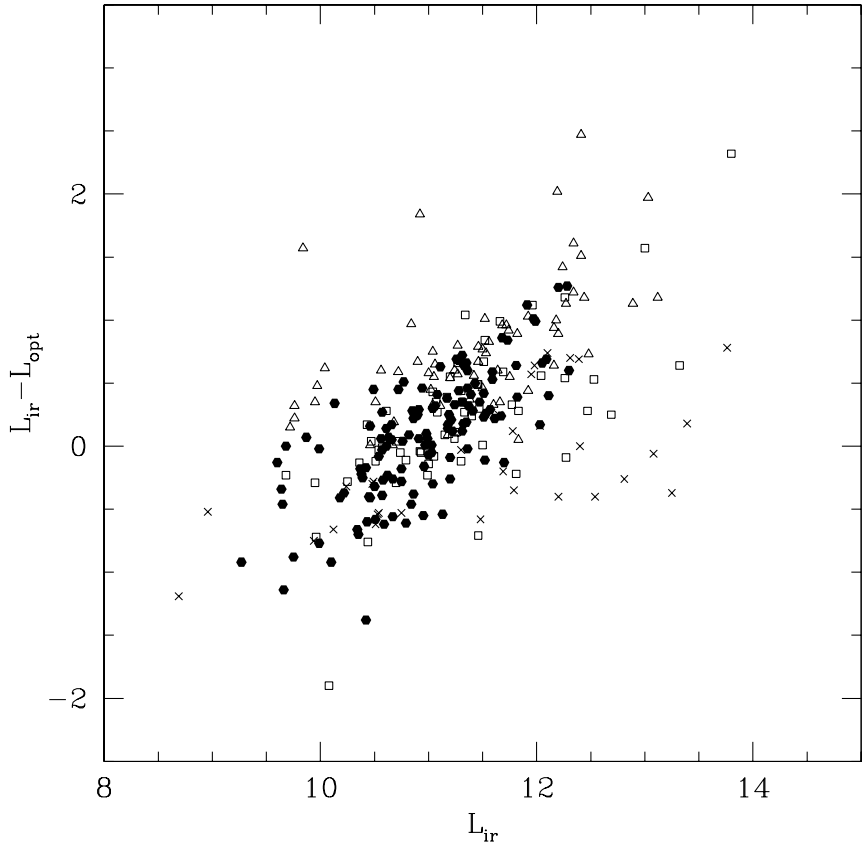,angle=0,width=12cm}
\caption{ $L_{ir} - L_{opt}$ versus $L_{ir}$ for ELAIS galaxies detected in 2 or more
ISO bands and with spectroscopic or photometric redshifts.  Filled circles: cirrus
sed, open squares: M82 starburst, open triangles: Arp 220 starbursts, crosses:
M82 starburst+AGN dust torus.
}
\end{figure*}

\begin{figure}
\epsfig{file=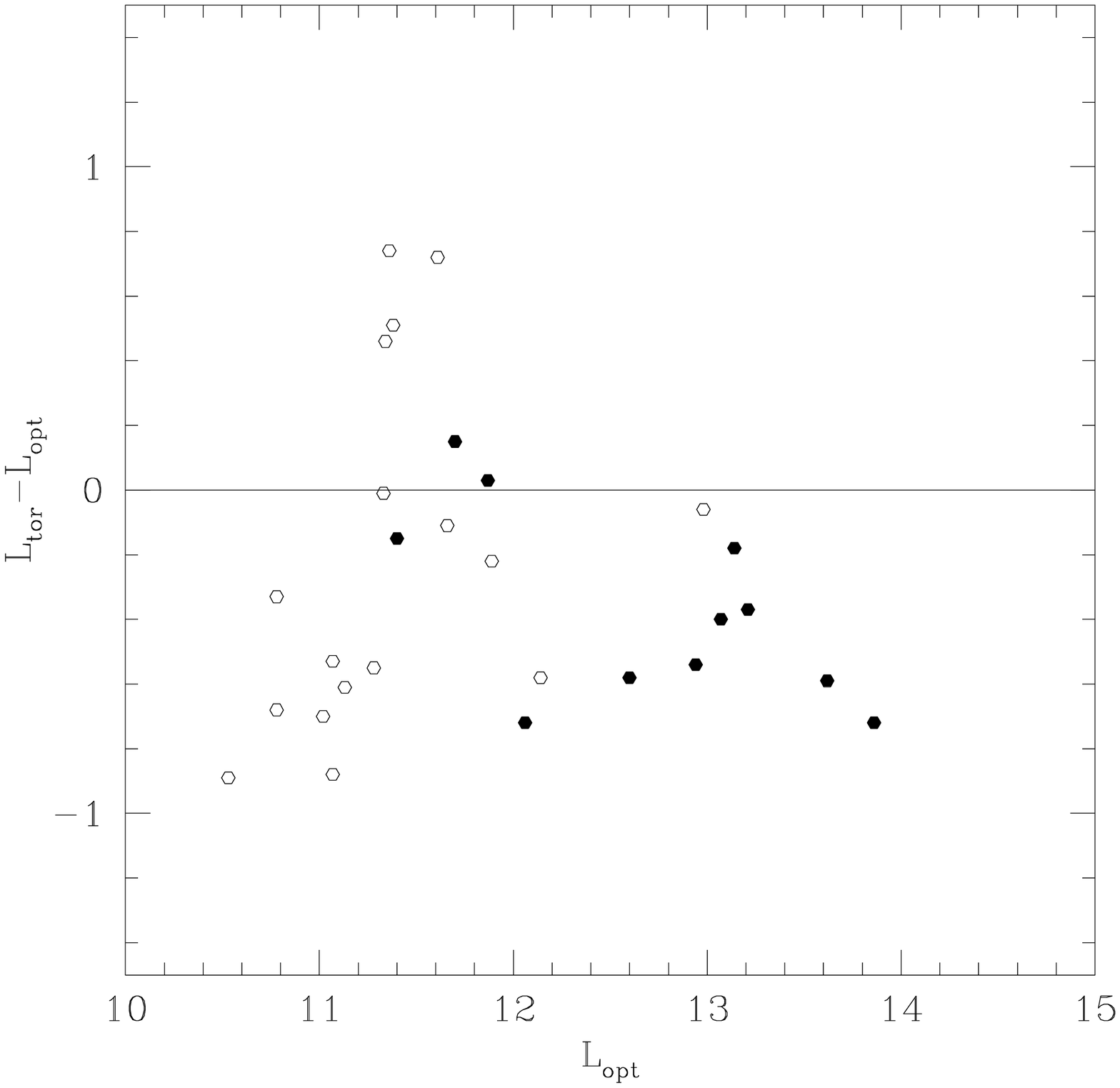,angle=0,width=8cm}
\caption{ $L_{ir} - L_{opt}$ versus $L_{ir}$ for ELAIS AGN dust tori detected in 2 or more
ISO bands and with spectroscopic or photometric redshifts.  Filled circles: Sy 1 (from optical spectroscopy)
, open circles: Sy 2, or
galaxy sed in optical.
}
\end{figure}

Galaxies with lower $L_{ir}$ and $L_{ir}-L_{opt} < 0.0$ are predominently fitted
with cirrus models, but cirrus models are also required at higher luminosities
and values of $L_{ir}-L_{opt}$ as high as 1.0.  This would seem to be consistent with 
the concept discussed by Efstathiou and Rowan-Robinson (2003) of cirrus galaxies
with $1 < A_V < 3$ and high total luminosities, as an alternative explanation of many
of the galaxies detected in submm surveys.  It is also consistent with the source-count
models of Rowan-Robinson (2001), in which the quiescent, cirrus galaxy component
undergoes the same strong luminosity evolution as starburst galaxies.
This high incidence of cold, very luminous, galaxies with high infrared-to-optical ratios is one of the main
new results of the ELAIS survey.
The expected trend towards M82 and Arp220 -type infrared seds as we go to higher infrared luminosities
and higher values of $L_{ir}-L_{opt}$ is also clearly seen.

Figure 31 shows a similar plot for galaxies for which an AGN dust torus component
is preferred.  Where galaxies are detected only at 6.7 and 15 $\mu$m, and starburst
luminosities in the hyperluminous range are implied, we have always prefered a
dust torus model as the more conservative assumption.  For Type 1 AGN, $L_{ir}-L_{opt}$
can be interpreted as the log of the covering factor.  The observed values are in the
range -0.7 to 0, implying covering factors in the range 20-100 $\%$, with a median value
of 30 $\%$.
Values greater than 0 would imply that the optical QSO must have suffered some extinction.

\section{Ultraluminous and hyperluminous infrared galaxies}

We have found a surprisingly large population of Arp220-like ultraluminous
infrared galaxies ($L_{ir} > 12.22 $) in this survey (14 $\%$ of the 15 $\mu$m galaxies with known z).  Morel
et al (2001) reported the first hyperluminous infrared galaxy in the survey.
In fact there are a total of 9 candidate hyperluminous galaxies ($L_{ir} > 13.22 $, Rowan-Robinson 2000)  in the survey,
listed in Table 4 (note that for galaxies ir sed-type = 6, the luminosity in the starburst and AGN dust torus components 
must be added to get the total infrared luminosity).  A further 45 galaxies would be hyperluminous if their 15 $\mu$m emission
were interpreted as due to a starburst sed (rather than the more conservative assumption
that it is due to AGN dust torus emission).
All 9 objects in Table 4 appear to be quasars: selection at 15 $\mu$m does favour detection of AGN dust tori.
For 4 galaxies this classification is based only on a photometric
redshift, and this needs to be confirmed by spectroscopy (especially where $z_{ph}$ = 2-2.5, since there is a
strong possibility of aliassing - see section 5).  The large number
of ultraluminous and hyperluminous infrared galaxies is probably a reflection of the very steep rise in the star-formation
rate between z = 0 and z = 2, and the associated strong evolution in the AGN
population (cf Pozzi et al 2003, LaFranca et al 2003), though gravitational lensing could also
play a part.

We also find 9 ERO's in the survey, defined as (r-K) $>$ 6, all from the K-band photometry of Vaisanen et al (2002) 
and Rigopoulou et al (2003, in preparation), and these are listed in Table 5.
3 of the objects have photometric redshifts and have the seds of elliptical galaxies at a
redshift $\sim$ 1.
Predicted (r-K) colours as a function of z show that elliptical galaxies have values $>$ 6 for $1 < z < 4$,

In the optical and near infrared the seds of dusty starbursts like M82 and Arp 220 look very like ellipticals due to the
extinction of the young stellar component.  Table 5 gives the values of $log_{10} (S_{15}/S_r)$ and $log_{10} (S_{rad}/S_r)$,
which are consistent with all 9 objects being highly extinguished starbursts like Arp 220, at  $z \sim 1$
(cf Arp 220 model curves in Figs 16 and 17).  However only a small fraction of EROs will fall into this category of
highly extinguished starbursts.  None of the 17 EROs with (R-K) $>$ 6 in Table 2 of Roche et al (2002), which are
located within a small 81.5 sq arcmin area of N2, are detected by ISO (one is an ELAIS radio source, 163657+410021).  
Counts of EROs in the N1 and N2 areas, and estimates of their space-density, are given by Vaisanen and Johansson (2003).
They conclude that the redshifts of their ERO sample lie in the range 0-7-1.5 and estimate the fraction of strong 
starbursts ( $> 30 M_o/yr$) to be $<$ 10$\%$.

\begin{table*}
\caption{Hyperluminous infrared galaxies in the ELAIS Catalogue}
\begin{tabular}{lllllllllllllll}
name & $m_r$ & $f_{WFS}$ & $n_{typ}$ & $z_{ph}$  & $z_{spect}$ & $n_{ztyp}$ & $n_{zref}$ & 
$L_{opt}$ & $n_{ISO}$ & $n_{irtyp}$ & $L_{ir}$ & $L_{tor}$\\
& & & & & & & & & & & & & & \\
ELAISC15-J002925-434917    &      18.63 &  -   & - & -   & 3.094 & 5 & 3 & 13.59 & 1 & 2 & 14.80 & 13.54\\
ELAISC15-J003213-434553    &      17.09 &  -   & - & -   & 1.707 & 5 & 3 & 13.57 & 1 & 2 & 13.97 & 13.29\\
ELAISC15-J050152-303519  &      17.86 & - & 4 & 0.0233  & 1.813 & 5 & 4 & 13.62 & 2 & 4 & 13.25 & 13.07\\
ELAISC15-J160419.0+541524  &      18.00 & 0.99 & 8 & 2.548  & - & - & - & 13.42 & 1 & 2 & 14.51 & 13.39\\
ELAISR 160758+542353        &      22.70 & 0.10 & 8 & 2.311 & - & - & - & 11.48 & 2 & 2 & 13.80 &-12.57\\
ELAISC15-J161259.2+541505  &      19.06 & 0.98 & 8 & 2.548  & - & - & - & 12.98 & 2 & 6 & 13.76 & 13.03\\
ELAISC15-J163739.2+405643  &      -     &  -   & - & -  & 1.438 & 5 & 1 &  - & 3 & 2 & 13.42 & 12.67\\
ELAISC15-J164010.1+410521  &      16.95 & 1.00 & 8 & 1.399  & 1.0990 & 5 & 6 & 13.17 & 3 & 6 & 13.39 & 12.92\\
ELAISC15-J164018.4+405812  &      18.06 & 0.99 & 8 & 2.311  & - & - & - & 13.33 & 1 & 2 & 14.33 & 13.30\\
\end{tabular}
\end{table*}

\begin{table*}
\caption{EROs in the ELAIS Catalogue}
\begin{tabular}{lllllll}
name & r' & K & $n_{typ}$ & $z_{phot}$ & $lg(S_{15}/S_r)$ & $lg(S_{rad}/S_r)$ \\
& & & & & & \\
ELAISR160721+544757 & 24.11 & 18.076 & - & - & - & 3.60 \\
ELAISC15-J160913.2+542320 & 24.29 & 17.180 & - & - & 3.48 & - \\
ELAISR161030+540247 & 23.53 & 16.400 & - & - & - & 2.91 \\
ELAISR161046+542329 & 23.15 & 17.090 & 2 & 1.138 & - & 2.28 \\
ELAISC15-J163536.6+404754 & 23.38 & 17.340 & 1 & 0.995 & 2.71 & - \\
ELAISR163555+412233 & 23.29 & 16.789 &  1 & 1.291 & 2.62 & - \\
ELAISR163723+410526 & 23.71 & 17.449 & - & - & - & 3.33 \\
ELAISC15-J163748.1+412100 & 24.50 & 17.926 & - & - & 3.14 & -  \\
ELAISR163758+411741 & 23.79 & 17.413 & - & - & - & 2.32\\
\end{tabular}
\end{table*}

\section{Discussion and conclusions}

We have presented the Final ELAIS Catalogue at U,g',r',i',Z, J,H,K, 6.7, 15, 90 and 175 $\mu$m, and 20 cm.
The process of band-merging and optical association of the sources has given considerable insight into the
extragalactic infrared populations present.  Although the largest single population is relatively nearby
(z $<$ 0.2) cirrus galaxies, there is a surprisingly large population of ultraluminous infrared galaxies
(14 $\%$ of 15 $\mu$m galaxies), many of them highly obscured starbursts like Arp 220.  There is also
a significant population of luminous ($L_{ir} > 11.5$) cool galaxies, consistent with the idea that the
quiescent component of star-formation in galaxies undergoes the same strong evolution as the starburst
component.  There appear to be a small proportion of
genuinely optically blank fields to r' = 24 (10 $\%$ at 15 $\mu$m, 3 $\%$ at 6.7 $\mu$m, up to 20 $\%$ at 90 $\mu$m,
and 1 $\%$ at 175 $\mu$m)
which must have high infrared-to-optical
ratios, be at higher redshift (z $>$ 0.6), and have high infrared luminosities.  They are therefore dusty
luminous starbursts or Type 2 AGN.
 
The ELAIS survey provides a strong basis for the SIRTF surveys.  The SWIRE Legacy Survey will include the 
ELAIS N1, N2, and S1 areas.  It will be worthwhile to obtain better seds and ir spectroscopy for the luminous 
infrared galaxies in the survey.  As SIRTF does not have a 15 $\mu$m band, the ELAIS data will be a useful 
complement to the SWIRE data.  The results presented here illustrate the importance of multiband optical 
data for photometric redshift determination.  The far infrared colour-colour diagrams discussed here 
illustrate the difficulties of trying to
determine redshifts from far infrared data alone.

\section{Acknowledgements}

The ELAIS consortium acknowledges support from EC TMR Networks 'ISO Survey'
and 'POE', and from PPARC.  DMA and OA acknowledge support from the Royal Society.
We thank the referee for helpful suggestions which have improved this paper.

\end{document}